\documentclass{aa}
\usepackage[varg]{txfonts}
\usepackage{caption-patch}
\usepackage{subfig}
\usepackage{textcomp}
\usepackage{graphicx,natbib,bm,gensymb}
\usepackage[colorlinks,citecolor=blue,linkcolor=blue]{hyperref}
\usepackage{stfloats}

\graphicspath{{./}{figures/}}
\bibpunct[, ]{(}{)}{;}{a}{,}{,}

\newcommand{\revised}[1]{#1}
\newcommand{\revsecond}[1]{{#1}}
\newcommand{\revthird}[1]{{{#1}}}

\makeatletter
\newcommand{\bianca}{\renewcommand\NAT@open{[}\renewcommand\NAT@close{]}}
\makeatother

\newcommand*\citesq[1]{{\bianca\cite{#1}}}

\newcommand\refsq[1]{[\ref{#1}]}

\let\oldincludegraphics\includegraphics%
\IfFileExists{.imgs_lowres/active}{%
\newcommand{\imgsprefix}{.imgs_lowres}%
}{%
\newcommand{\imgsprefix}{.}
}
\renewcommand{\includegraphics}[2][]{\oldincludegraphics[#1]{\imgsprefix/#2}}%

\usepackage[normalem]{ulem}
\usepackage{authorreview}
\addreviewauthor{thomas}{TR}{orange}

\begin{document}

\title{Modeling the nonaxisymmetric structure in the HD~163296 disk with
planet-disk interaction}

\author{P. J. Rodenkirch\inst{1} \and
Thomas Rometsch \inst{2} \and
C. P. Dullemond \inst{1} \and
Philipp Weber \inst{3} \inst{4} \inst{5} \and
Wilhelm Kley \inst{2}}
\institute{Institute for Theoretical Astrophysics, Zentrum f\"ur Astronomie, Heidelberg University, Albert-Ueberle-Str. 2, 69120 Heidelberg, Germany \and
Institut für Astronomie und Astrophysik, Universit\"at T\"ubingen,
  Auf der Morgenstelle 10, 72076 T\"ubingen, Germany \and
  Niels Bohr International Academy, The Niels Bohr Institute, University of Copenhagen, Blegdamsvej 17, DK-2100 Copenhagen Ø, Denmark
  \and 
  Departamento de Astronom\'{\i}a, Universidad de Chile, Casilla 36-D, Santiago, Chile
  \and
  Departamento de F\'{\i}sica, Universidad de Santiago de Chile, Av. Ecuador 3493, Estaci\'{o}n Central, Santiago, Chile
  }

\date{\today}
 
\abstract{High resolution ALMA observations like the DSHARP campaign revealed
a variety of rich substructures in numerous protoplanetary disks. These
structures consist of rings, gaps and asymmetric features. It is debated
whether planets can be accounted for these substructures in the dust
continuum. Characterizing the origin of asymmetries as seen in HD~163296
might lead to a better understanding of planet formation and the underlying
physical parameters of the system.} {We test the possibility of the formation
of the crescent-shaped asymmetry in the HD~163296 disk through planet disk
interaction. The goal is to obtain constraints on planet masses and
eccentricities and disk viscosities.\revised{We furthermore test the
reproducibility of the two prominent rings in the HD~163296 disk at 67~au and
100~au.}} {Two dimensional, multi-fluid, hydrodynamical simulations are
performed with the FARGO3D code including three embedded planets. Dust is
described with the pressureless fluid approach and is distributed over eight
size bins. Resulting grids are post-processed with the radiative transfer
code RADMC-3D and the CASA software to model synthetic observations.} {We
find that \revthird{the crescent-shaped asymmetry} can be qualitatively modeled with a Jupiter mass
planet at a radial distance of 48~au. Dust is trapped preferably in the
trailing Lagrange point L5 with a mass of 10 to 15 earth masses. The
observation of such a feature confines the level of viscosity and planetary
mass. Increased values of eccentricity of the innermost Jupiter mass planet
damages the stability of the crescent-shaped feature and does not reproduce the
observed radial proximity to the first prominent ring in the system.
Generally, a low level of viscosity ($\alpha \leq 2\cdot10^{-3}$) is
necessary to allow the existence of such a feature.\\ Including dust
feedback the leading point L4 can dominantly capture dust for dust grains
with an initial Stokes number $\leq 3.6\cdot 10^{-2}$. \revised{In the synthetic ALMA
observation of the model with dust feedback two crescent-shaped features are
visible. The observational results suggest a negligible effect of dust
feedback since only one such feature has been detected so far. The dust-to-gas
ratio may thus be overestimated in the models. Additionally, the planet mass
growth time scale does not strongly affect the formation of such asymmetries
in the co-orbital region.}}{}

\keywords{protoplanetary disks - planet-disk interactions - planets and
satellites: formation - planets and satellites: rings - hydrodynamics -
radiative transfer}

\titlerunning{nonaxisymmetric structures in HD 163296}

\maketitle

\section{Introduction}
\label{sec:intro}
In the advent of high angular resolution millimeter continuum observations
with the Atacama Large Millimeter and Submillimeter Array (ALMA) insights
into substructures of protoplanetary disks have become available. The
striking results of the first highly resolved observation of the
protoplanetary disk HL Tau unveiled a rich variety of concentric rings in the
dust continuum \citep{alma_hltau_2015}.
An extensive survey with the goal to image detailed structures in 20 disks
was performed with the DSHARP campaign \citep{Andrews2018} with unprecedented
resolution. Rings and gaps seem to be ubiquitous in these disks and appear
independently of the stellar luminosity \citep{Huang2018a}. A subset of the
observations display nonaxisymmetric structures like spirals and
crescent-shaped features. \\ Currently, it is debated whether these
structures are signposts of embedded planets \citep{Zhang2018}. Planet-disk
interaction has been a central topic in the dynamics of protoplanetary disks,
first with analytic studies of resonances and spiral density waves
\citep{goldreich_tremaine_1979, goldreich_tremaine_1980} or planetary
migration \citep{lin_papaloizou_1986}.\\ Jupiter mass planets can open a gap
in the gas \citep{kley_1999}. As shown by numerical, two fluid simulations by
\cite{paardekooper_mellema_2004} a planet of 0.1 Jupiter masses
($M_\mathrm{jup}$) is sufficient to open a gap in the dust. Even lower masses
down to 0.05~$M_\mathrm{jup}$ can lead to gap formation if mainly mm-sized
dust particles are present \citep{paardekooper_mellema_2006}. Dust structures
created by planet-disk interaction are generally more diverse than their
counterpart in the gas \citep{fouchet_2007,maddison_2007}. For massive
planets of 5~$M_\mathrm{jup}$ and cm-sized grains \cite{fouchet_2010} found
azimuthally asymmetric dust trapping in the context of 3D SPH simulations.
Embedded dust grains are prone to drift towards pressure maxima in the disk
\citep{Whipple1972}. Thus, perturbations caused by a sufficiently massive
planet can efficiently trap up to meter-sized bodies on the outer edge of the
gap, form a ring structure and may aid planetesimal formation
\citep{ayliffe_2012}.\\ \cite{dong_gaps_2015} found that multiple planets can
explain large cavities at near-infrared and millimeter wavelengths as
observed in transition disks \citep{calvet_transition_2005,
hughes_transition_2009}. In a system with two planets dust trapped in the
leading and trailing Lagrange points (L4 \& L5) can be a transient feature,
depending on the outer planet \citep{picogna_2015}. In the context of low
viscosity disks multiple rings and gaps emerge with a single planet through
shocks of the primary and secondary spiral arm \citep{zhu_2014,
bae_rings_2017}.\\ Torques caused by the gravitational interaction between
planets and the disk lead to migration (\citesq{kley_nelson_2012} for a
review) which in turn affects the observable dust substructures, e.g. changes
in the ring intensity or asymmetric triple ring structures depending on the
migration rate and direction \citep{meru_2018, Weber2019}. Migration is
sensitive to the underlying disk physics and can be chaotic in very low
viscosity disks \citep{mcnally_2019}. \revised{In general, protoplanetary disks seem
to be only weakly turbulent, indicating regimes of $\alpha$ on the order of
$10^{-4}$ to $10^{-3}$\citep{flaherty_2015, flaherty_2017, Dullemond2018},
using the turbulent $\alpha$ viscosity parametrization from
\cite{Shakura1973}. When referring to low viscosity in disks, the magnitude
of the effective $\alpha$ viscosity is meant which drives angular momentum
transport and thus accretion due the underlying turbulent processes. Hence, a
low effective viscosity can be linked to weak turbulence.}\\
Spiral waves in the gas, excited by a planet, are mostly hidden in the dust
dynamics, favoring gaps and rings \citep{dipierro_2015}. The gravitational
instability \citep{toomre_1964} in sufficiently massive disks can however
trigger spiral waves trapping large particles \citep{rice_2004, rice_2006}.
These waves are in principle also observable in scattered light observations
\citep{pohl_2015}.\\ Nonaxisymmetric features like vortices can be created
by the Rossby wave instability \citep{lovelace_1999, li_rossby_2000} enabling
dust trapping \citep{baruteau_vortex_2016}. Observationally these might be
visible as "blobs" or crescent-shaped features as seen in IRS 48
\cite{vandermarel_irs48_2013} or HD 135344B \cite{cazzoletti_vortex_2018}.
Alternatively, hydrodynamic instabilities like the baroclinic instability
\citep{klahr_baroclinic_2003} or the vertical shear instability are able to
form vortices \citep{manger_2018}. \\ In the presence of weak magnetic fields
the magneto-rotational instability (MRI) triggers turbulence
\citep{balbus_1991} and drives accretion flows. If the disk mid plane is
effectively shielded from ionizing radiation the inner part of the disk
becomes laminar, the so-called dead zone, with layered accretion on the
surface level \citep{gammie_1996}. In the outer parts of the disk high energy
photons may ionize the gas sufficiently to activate the MRI. The transition
between the dead zone and the MRI-active region and thus the change in
turbulent viscosity can also create ring structures \citep{flock_2015}.\\
With all these possible substructure formation mechanisms at hand, it is of
interest to identify markers of the presence of planets embedded in
protoplanetary disks. A popular and well-studied disk is the one around the
Herbig Ae star HD~163296 at a distance of 101 pc \citep{gaia2018}. The
appearance in the 1.25~mm continuum emission of the disk is dominated by two,
already beforehand observed rings at a radial distance of 67~au and 100~au
relative to the central star respectively \citep{Isella2016, Isella2018}. An
additional faint ring has been detected at 159~au. An intriguing feature is a
crescent-shaped asymmetry within the inner gap located at 48~au
\citep{Huang2018a}. The feature itself is situated at a radial distance of
55~au, thus with an offset of 7~au from the gap center \citep{Isella2018}. An
image of the original observation is show in Fig.~\ref{fig:synth_images}.\\
\begin{figure}[ht] 
  \centering
  \includegraphics[width=\linewidth]{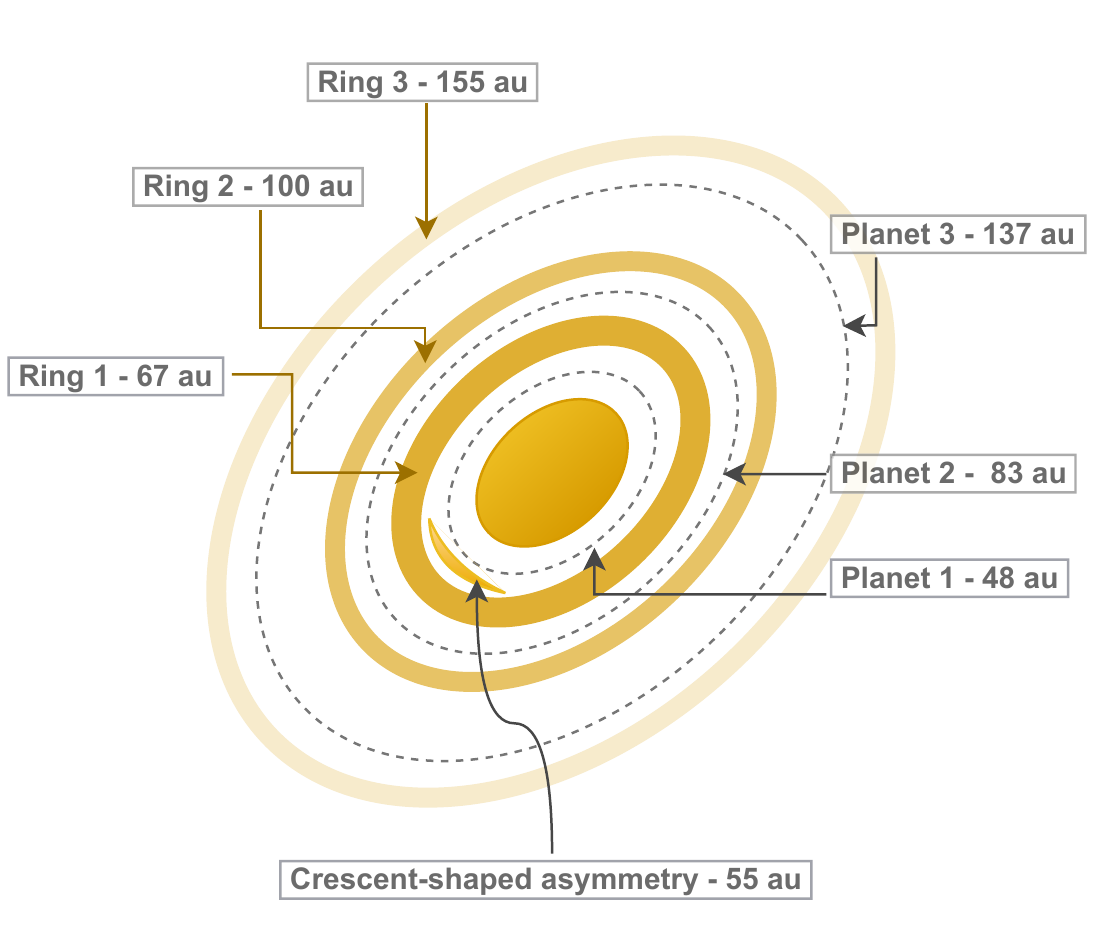}
  \caption{\revised{Sketch of the oberserved dust substructure of the HD~163296
  system. The colored rings and the crescent-shaped feature mimic the dust
  while the grey dashed lines indicate the semi-major axes of the modeled
  planetary system.
  The labels introduced in this figure will be used throughout the text.}}
  \label{fig:disk_sketch}
\end{figure}
The origin of such a structure is unknown and a preliminary model was
presented in \cite{Zhang2018} involving planet-disk interaction. In these
models asymmetries in the co-orbital region are common if the viscosity is
low.\\ Before the publication of the results from the DSHARP campaign, the
HD~163296 disk was modeled by \cite{Liu2018}. Their models incorporated 2D
two-fluid hydrodynamical simulations with three planets in their respective
positions matching the observed gaps. With synthetic images using radiative
transfer calculations they could match the observed density profile with
0.46, 0.46, and 0.58 Jupiter masses for the three planets and a radially
increasing turbulent viscosity parametrization. \\ In the suite of
simulations by \cite{Zhang2018} using hydrodynamical models with Lagrangian
particles the proposed mass fits are 0.71, 2.18, and 0.14 Jupiter masses for
an $\alpha$-viscosity of $10^{-3}$. In their lower viscosity models of
$\alpha = 10^{-4}$ masses of 0.35, 1.07, and 0.07 Jupiter masses were fitted.
\\ Further observational constraints of the hypothetical two outer planets
were provided by kinematical detections by \cite{Teague2018}. Their model
predicts masses of 1 and 1.3 Jupiter masses for these planets.
\cite{pinte_2020} argue that velocity "kinks" observed in the CO observations
with ALMA are evidence of nine planets in the DSHARP sample, including two
planets in the HD~163296 disk at 86~au and 260~au. The signal-to-noise ratio
is not sufficient to probe the inner gap at 48~au. \\ In this paper we want
to further explore the possibility of reproducing the observed structures by
planet-disk interaction with a focus on \revthird{the crescent-shaped asymmetry} in the dust
emission. \revthird{This asymmetric feature} has been present in the works discussed above
but it has not been subject to more detailed analysis yet. Given the
motivation of the crescent-shaped feature in the observation of HD~163296 we
aim to constrain the visibility of such an agglomeration of dust caused by
planet-disk interaction and its dependence on the physical parameters of the
system like planet mass, turbulent viscosity and dust size. In comparison to
the study of \cite{Zhang2018} we employ two-dimensional hydrodynamical models
with a fluid formulation of dust.\\ Sec.~\ref{sec:model} introduces the
physical model and code setup as well as the post processing pipeline to
predict observable features. In Sec.~\ref{sec:results} we present the main
results of our study. Sec.~\ref{sec:discussion} compares these findings with
previous works and addresses limitations of the model. In
Sec.~\ref{sec:conclusion} we summarize the main results with concluding
remarks.

\section{Model} 
\label{sec:model}

All hydrodynamical models presented in this work were performed with the
FARGO3D multi-fluid code \citep{Benitez-llambay2016, Benitez-llambay2019}
making use of an orbital advection algorithm \citep{masset_fargo_2000}.
\revised{
The code is based on the public version of FARGO3D with the addition of
allowing a constant dust size throughout the simulation and a spatially
variable viscosity.}

\subsection{Basic equations}
The FARGO3D code solves the conservation of mass
(Eqs.~\refsq{eq:gas_cons_mass} and~\refsq{eq:dust_cons_mass}) and
conservation of momentum (Eqs.~\refsq{eq:gas_cons_mom}
and~\refsq{eq:dust_cons_mom}) for gas and dust in our model setups:

\begin{flalign}
\label{eq:gas_cons_mass}\partial_\mathrm{t} \Sigma_\mathrm{g} + \nabla \cdot \left( \Sigma_\mathrm{g} \, \bm{v}_\mathrm{g} \right) = 0 \\
\label{eq:gas_cons_mom}\Sigma_\mathrm{g} \left[ \partial_\mathrm{t} \bm{v}_\mathrm{g} +  \left( \bm{v}_\mathrm{g} \cdot \nabla \right) \bm{v}_\mathrm{g} \right] = - \nabla P + \nabla \cdot \Pi - \Sigma_\mathrm{g}\nabla \Phi - \sum_\mathrm{i} \Sigma_\mathrm{i} \mathrm{f}_\mathrm{i} \\
\label{eq:dust_cons_mass}\partial_\mathrm{t} \Sigma_\mathrm{d, i} + \nabla \cdot \left( \Sigma_\mathrm{d, i} \, \bm{v}_\mathrm{d, i} + \bm{j}_\mathrm{i} \right) = 0  \\
\label{eq:dust_cons_mom}\Sigma_\mathrm{d, i} \left[ \partial_\mathrm{t} \bm{v}_\mathrm{d, i} +  \left( \bm{v}_\mathrm{d, i} \cdot \nabla \right) \bm{v}_\mathrm{d, i} \right] = - \Sigma_\mathrm{d, i}\nabla \Phi + \Sigma_\mathrm{d, i} \mathrm{f}_\mathrm{i}
\end{flalign}
Here, $\Sigma_\mathrm{g}$ denotes the gas surface density,
$\Sigma_{\mathrm{d}, \mathrm{i}}$ the corresponding dust species, $P =
\Sigma_\mathrm{g} c_\mathrm{s}^2$ the gas pressure, linked to the density by
a locally isothermal equation of state with the sound speed $c_\mathrm{s}$,
$\Phi$ the gravitational potential of the star and planets, $\Pi$ the viscous
stress tensor, $f_\mathrm{i}$ the interaction forces between gas and dust and
$\bm{v}_\mathrm{g}$ and $\bm{v}_{\mathrm{d}, i}$ the gas and dust velocities
respectively. Dust feedback is included by the term $\sum_\mathrm{i}
\Sigma_\mathrm{i} f_\mathrm{i}$ in Eq.~\ref{eq:gas_cons_mom}.\\ We consider
turbulent mixing and diffusion of dust grains by using the dust diffusion
implementation described in \cite{Weber2019}. The corresponding diffusion
flux $\bm{j}_\mathrm{i}$ can be written as
\begin{equation}
\bm{j}_\mathrm{i} = - D_\mathrm{i} \left( \Sigma_\mathrm{g} + \Sigma_{\mathrm{d}, \mathrm{i}} \right) \nabla \left( \frac{\Sigma_{\mathrm{d}, \mathrm{i}}}{\Sigma_\mathrm{g} +  \Sigma_{\mathrm{d}, \mathrm{i}}} \right) ,
\end{equation}
with the diffusion constant $D_\mathrm{i}$ being proportional to the turbulent viscosity $\nu$ \citep{Youdin_Lithwick2007}:
\begin{equation}
D_\mathrm{i} = \nu \frac{1 + \mathrm{St}_\mathrm{i}^2}{\left( 1 + \mathrm{St}_\mathrm{i}^2 \right)^2} .
\end{equation}
The Stokes number $\mathrm{St}_\mathrm{i}$ of dust species $\mathrm{i}$ is proportional to the stopping time $t_\mathrm{stop}$ and can be written as
\begin{equation} 
\label{eq:stokes_number}
\mathrm{St}_\mathrm{i} = t_\mathrm{stop} \, \Omega_\mathrm{K} = \frac{\pi}{2} \frac{a_\mathrm{i} \, \rho_\mathrm{d}}{\Sigma_\mathrm{g}} ,
\end{equation}
where $\Omega_\mathrm{K}$ is the Keplerian angular frequency, $a_\mathrm{i}$ the dust grain size and $\rho_\mathrm{d}$ the material density of the grains.
The gas-dust interaction is modeled using the Epstein drag law, which is expected to be valid if the particle size is smaller than the mean-free path of surrounding gas molecules. Here, the drag force is proportional to the relative velocities between the corresponding fluids \citep{Whipple1972}.  The drag force $\bm{f}_\mathrm{i}$ can be expressed as
\begin{equation}
\bm{f}_\mathrm{i} = - \frac{\Omega_\mathrm{K}}{\mathrm{St}_\mathrm{i}} \left( \bm{v}_{\mathrm{d}, \mathrm{i}} - \bm{v}_\mathrm{g} \right) .
\end{equation} 

\subsection{Disk model}
In the disk model the planet-disk interaction is implemented as an additional
smoothed potential term (Plummer-potential) for each planet. The smoothing
length is set to $0.6 H$, where $H(r) = c_\mathrm{s}(r) /
\Omega_\mathrm{K}(r)$ is the pressure scale height at the radial distance
$r$. The specific factor acts as a correction for 3D effects in the 2D
simulation \citep{Mueller2012}. \\ 
Three planets are modeled in the simulations. The two
outer ones are set to the locations indicated by \cite{Teague2018}. The inner
planet is put at the corresponding gap location while the mass is varied in
the different runs. In the fiducial models, the planet locations and masses
are $r_\mathrm{p} = \{ 48, 83, 137\}$~au, $M_\mathrm{p0} = \{ 1.0, 0.55, 1.0
\}\, M_\mathrm{jup}$, respectively. From this point on we will refer to the
three planets as planet~1, 2 and 3. The same notation will be used for the
apparent ring structures, i.e. ring~1: observed or modeled ring at 67~au;
ring~2 at 100~au (see Fig.~\ref{fig:disk_sketch}). \revised{We chose lower
the mass of planet~2 compared to the one predicted in \cite{Teague2018} since
it allows a sufficiently massive ring~2 while not significantly disturbing
the \revised{crescent-shaped asymmetry} by its repeated gravitational interaction.\\
For the fiducial model the corresponding parameter variations we set the planet 
mass $M_\mathrm{p}$ to its final value $M_\mathrm{p0}$ at the beginning of the simulation.
The mass growth time scale however is known to have an impact on the
formation of disk structures, such as vortices
\citep{hammer_2019,hallam_2020}. We therefore investigate the robustness of
the results by testing different planet growth time scales:
\begin{equation}
  M_\mathrm{p}(t) = \frac{1}{2} \left[ 1 - \mathrm{cos}\left( \pi \frac{t}{T_\mathrm{G}} \right) \right] \, M_\mathrm{p0}
\end{equation}
$T_G$ refers to the planet growth time scale ranging from $10 T_0$ to $500
T_0$ with $T_0$ denoting the orbital period at $r_0 = 48~\mathrm{au}$. 
By using this simplified growth prescription we do not directly model the accretion of
gas onto the planets and we thus do not artificially remove any mass from the
simulation domain.
Furthermore, for all models the displacement of the center of mass by the
influence of the planets is taken into account as an additional indirect term
added to the potential.
The semi-major axes of the planets are kept fixed throughout the whole simulation. \\
}
The initial surface density profile is
assumed to be a power law with an exponential cutoff: 
\begin{equation}
\Sigma_{\mathrm{g} / \mathrm{d}} = \Sigma_{\mathrm{g}/\mathrm{d}, 0} \left(
\frac{r}{r_0} \right)^{-\mathrm{p}} \mathrm{exp}\left[ -\left( \frac{r}{r_0}
\right)^\mathrm{s} \right] \,, 
\end{equation} 
with $r_0 = 48 \,\text{au}$ and
the initial gas and dust surface densities $\Sigma_{\mathrm{g},0}$ and
$\Sigma_{\mathrm{d},0}$. Similar to the models of \cite{Liu2018} we choose a
surface density slope of $\mathrm{p} = 0.8$. For the dust a sharper cutoff of
$\mathrm{s} = 2$ was chosen compared to the gas cutoff of $\mathrm{s} = 1$.
In all simulation runs an initial dust-to-gas ratio of $\left(\sum_\mathrm{i}
\Sigma_{\mathrm{d}, \mathrm{i}} \right) / \Sigma_\mathrm{g} = 0.01$ is
assumed.\\ 
We approximate the disk thermodynamics with a locally isothermal
equation of state. The model is parametrized through locally isothermal sound
speed 
\begin{equation} 
  c_\mathrm{s} = \frac{H(r_0)}{r_0} \left( \frac{r}{r_0}
\right)^{\frac{1}{4}} \, \Omega_\mathrm{K} r . 
\end{equation} 
This corresponds to a flared disk with flaring index 0.25. The aspect ratio
$H / r$ at $r_0$ is set to a value of 0.05. Assuming a mean molecular weight
of $\mu = 2.353$ the mid plane temperature profile can be written in the
following way 
\begin{equation} 
  T_\mathrm{mid} (r) = \sqrt{\frac{\mu
m_\mathrm{p}}{k_\mathrm{B}}} \, c_\mathrm{s} \approx 25 \, \left(
\frac{r}{r_0} \right)^{-\frac{1}{2}} \mathrm{K} , 
\end{equation} 
with the proton mass $m_\mathrm{p}$. The temperature at 48~au matches the
findings of \cite{Dullemond2019}. \\
We assume a radially smoothly increasing
turbulent viscosity profile, motivated by Eq. 4 in the work of \cite{Liu2018}
and similar dead zone parametrizations presented in \cite{pinilla_2016} and
\cite{miranda_2016}: 
\begin{equation} 
  \alpha(r) = \alpha_\mathrm{min} \left\{
1 - \frac{1}{2} \left( 1 - \frac{\alpha_\mathrm{max}}{\alpha_\mathrm{min}}
\right) \left[ 1 - \mathrm{tanh}\left( -\frac{r - R}{\sigma r_0} \right)
\right] \right\} , 
\end{equation} 
where the parameters $R$ and $\sigma$ are set to 144~au and 1.25
respectively. Similar to the values in \cite{Liu2018} the parameter $R$
refers to the mid-point of the transition in $\alpha$ whereas $\sigma$
defines the slope.\\ 

\begin{figure}[t] 
  \centering
  \includegraphics[width=\linewidth]{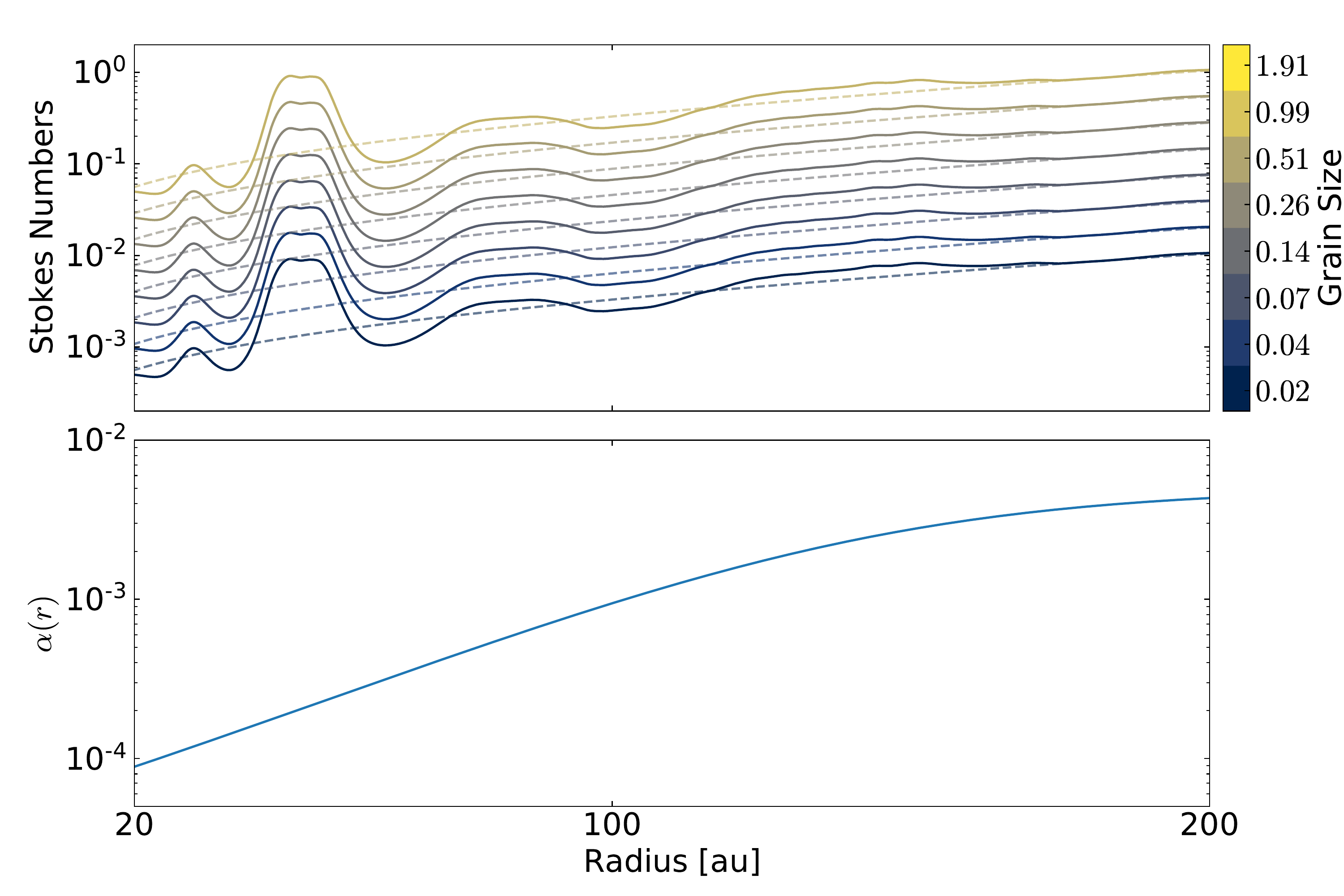}
  \caption{\textit{Upper panel:} \revised{azimuthally averaged values of the Stokes number St
  of model \texttt{fid}. The dashed lines indicate the initial values and the solid lines represent the state after $500\,T_0$. \\
  \textit{Lower panel:} prescribed $\alpha$ viscosity. The
  blue line visualizes the radially increasing $\alpha$ viscosity set in the
  model.}
  }
  \label{fig:stokes_numbers_alpha}
\end{figure}

\subsection{Boundary conditions}
The radial velocities in the ghost cells are set according to anti-symmetric
boundary conditions. On a staggered mesh this corresponds to
$v_\mathrm{r}(r_\mathrm{ghost}) = -v_\mathrm{r}(r_\mathrm{act})$, where
$r_\mathrm{ghost}$ is the ghost cell and $r_{act}$ the equivalent mirrored
cell on the active hydro mesh near the boundary. The value at the staggered
boundary itself is set to zero. The azimuthal velocities are set to the
initial Keplerian profile in the ghost zone.\\ 
The surface density is
extrapolated according to the density gradient exponent p. Additionally, wave
damping is applied within 15\% of the respective boundary radius. In this
region the density is exponentially relaxed towards the initial values within
0.3 local orbital time scales. The procedure follows the wave damping
boundary conditions used in \cite{devalborro_2006}. \\
\revised{
We tested the robustness of the wave damping with respect to the chosen
inner boundary condition. No significant wave reflections are detected for
both a symmetric and anti-symmetric inner boundary condition. The relative
difference between these two approaches is on the order of $10^{-5}$ compared
to the reflected perturbations of $\approx 10^{-2}$ without wave damping.
}
\subsection{Code setup and parameters}\label{sec:parameters}
Various ranges of individual disk parameters were considered for constraining
the impact onto dust features in the simulations. Table~\ref{tab:simulations}
gives an overview over the performed simulations and their parameter choices.
\\ For most of the runs a resolution of 560 radial and 895 azimuthal cells
was chosen where the grid is logarithmically spaced in radial direction.
Compared to the local disk scale height a ratio of roughly 7 cells per scale
height is achieved in each direction. \revised{
The models with the suffix \texttt{\_dres} were run with a resolution of 14 cells per scale height.
\revthird{
The low resolution runs are subject to a larger numerical diffusion compared to the high resolution models.
Although these runs can be used for mass estimates of the crescent-shaped feature, the results should be taken with caution concerning the dust substructure lifetime.
Whenever this difference becomes significant we default to the high resolution runs. Further details are given in appendix~\ref{sec:appendix_res_study}.
}
For all simulations the mass of the central star is set to $1.9 \,M_\odot$
}
The parameters $\alpha_\mathrm{min}$
and $\alpha_\mathrm{max}$ are set to $10^{-5}$ and $5 \cdot 10^{-3}$ which
results in a value of $\alpha(r_0) \approx 2 \cdot 10^{-4}$ at the location
of $r_0 = 48 \,\mathrm{au}$. \revsecond{The radial profile of $\alpha(r)$ is shown in the lower panel of Fig.~\ref{fig:stokes_numbers_alpha}.}
\\
The dust is sampled by 8 separate fluids with
Stokes numbers spaced logarithmically ranging from $1.3 \cdot 10^{-3}$ to
$1.3 \cdot 10^{-1}$ at $r_0 = 48 \, \mathrm{au}$. In the code, the equivalent
grain size at $r_0$ is applied to the whole domain and is kept constant
throughout the whole simulation. With an initial value of
$\Sigma_{\mathrm{g},0} = 37.4 \, \mathrm{g} / \mathrm{cm}^2$ the minimum and
maximum grain sizes are $a_\mathrm{min} = 0.19 \, \mathrm{mm}$ and
$a_\mathrm{max} = 19 \, \mathrm{mm}$. \revised{No dust size evolution is modeled here. It should be noted that these
state the initial values of St and changes with time depending on the gas
surface density, as shown in the upper panel of Fig.~\ref{fig:stokes_numbers_alpha}. The most prominent change of an increase of
about one order of magnitude in St occurs at the gap carved by planet~1 after 500 orbits.
}
 \\
The majority of models neglects dust feedback to the motion of the gas. 
Each dust species can thus be scaled in density individually without
violating the validity of the dynamical features.\\
\revised {
Per default, the simulations are executed until $1000 \, T_0$. Simulation runs
with a nonzero growth time scale $T_\mathrm{G}$ and the model \texttt{fid\_dres} are run until $2000 \, T_0$.
After $\approx 900$ orbits the gas and dust structure converges. The
\revthird{crescent-shaped asymmetry} builds up to a stable niveau after less than 100 orbits. We refer
to Sec. \ref{sec:results} for more details.
}
\begin{table*}[ht]
  \begin{center}
    \caption{Relevant simulation runs and their respective numerical parameters.}
	    \renewcommand\arraystretch{0.8}
    \label{tab:simulations}
    \begin{tabular}{lcccccccccc}
      \hline\hline                       \\[-0.6em]
      Simulation & H/r & $M_\mathrm{pl1}$ [$M_\mathrm{jup}$] & $M_\mathrm{pl2}$ [$M_\mathrm{jup}$] & $e$ & $N_r$ & $N_\phi$ & $r_\mathrm{cut}$ [au] & $\alpha$ & $T_\mathrm{G}$ \\[0.2em]
      \hline                             \\[-0.6em]
      \texttt{fid} & 0.05   & 1       & 0.55 &      & 560 & 895     &                   &  $2 \cdot 10^{-4}$ & 0  \\
      \texttt{fid\_dres} & 0.05   & 1       & 0.55 &      & 1120 & 1790     &                   &  $2 \cdot 10^{-4}$ &  \\
      \texttt{hr4}  & 0.04   &         &      &      &     &         &                  & &    \\
      \texttt{hr45} & 0.045  &         &      &      &     &       &                    & &    \\
      \texttt{hr55} & 0.055  &         &      &      &     &       &                    & &    \\
      \texttt{hr6}  & 0.06   &         &      &      &     &         &                  & &    \\
      [0.2em]\hline                                                                     & &    \\[-0.8em]
      \texttt{taper10} &  &        &  &      &  &   &                   &   & 10  \\
      \texttt{taper50} &  &        &  &      &  &   &                   &   & 50  \\
      \texttt{taper100} & &        &  &      &  &   &                   &   & 100  \\
      \texttt{taper500} & &        &  &      &  &   &                   &   & 500  \\
      [0.2em]\hline                                                                     & &    \\[-0.8em]
      \texttt{p1m1}  &        & 0.20    &      &      &     &         &                 & &    \\
      \texttt{p1m2}  &        & 0.36    &      &      &     &         &                 & &    \\
      \texttt{p1m3}  &        & 0.52    &      &      &     &         &                 & &    \\
      \texttt{p1m4}  &        & 0.68    &      &      &     &         &                 & &    \\
      \texttt{p1m5}  &        & 0.84    &      &      &     &         &                 & &    \\
      [0.2em]\hline                                                                     & &    \\[-0.8em]
      \texttt{p1m1fb} &        & 0.20    &      &      &     &         &                & &    \\
      \texttt{p1m2fb} &        & 0.36    &      &      &     &         &                & &    \\
      \texttt{p1m3fb} &        & 0.52    &      &      &     &         &                & &    \\
      \texttt{p1m4fb} &        & 0.68    &      &      &     &         &                & &    \\
      \texttt{p1m5fb} &        & 0.84    &      &      &     &         &                & &    \\
      \texttt{p1m6fb} &        &         &      &      &     &         &                & &    \\
      \texttt{p1m6fb\_dres} &        &         &      &      &  1120   &     1790    &  & &    \\
      [0.2em]\hline                                                                     & &    \\[-0.8em]
      \texttt{p2m1} &        &         & 0.30 &      &     &         &                  & &    \\
      \texttt{p2m2} &        &         & 0.42 &      &     &         &                  & &    \\
      \texttt{p2m3} &        &         & 0.54 &      &     &         &                  & &    \\
      \texttt{p2m4} &        &         & 0.66 &      &     &         &                  & &    \\
      \texttt{p2m5} &        &         & 0.78 &      &     &         &                  & &    \\
      \texttt{p2m6} &        &         & 0.90 &      &     &         &                  & &    \\
      [0.2em]\hline                                                                     & &    \\[-0.8em]
      \texttt{ecc1} &        &         &      & 0.02 &     &     &                      & &    \\
      \texttt{ecc2} &        &         &      & 0.04 &     &     &                      & &    \\
      \texttt{ecc3} &        &         &      & 0.06 &     &     &                      & &    \\
      \texttt{ecc4} &        &         &      & 0.08 &     &     &                      & &    \\
      \texttt{ecc5} &        &         &      & 0.10 &     &     &                      & &    \\
      [0.2em]\hline                                                                     & &    \\[-0.8em]
      \texttt{cut1} &        &         &      &      &     &     & 150                  & &    \\
      \texttt{cut2} &        &         &      &      &     &     & 175                  & &    \\
      \texttt{cut3} &        &         &      &      &     &     & 200                  & &    \\
      \texttt{cut4} &        &         &      &      &     &     & 225                  & &    \\
      \texttt{cut5} &        &         &      &      &     &     & 250                  & &    \\
      [0.2em]\hline                                                                     & &    \\[-0.8em]
      \texttt{alpha1} &        &         &      &      &     &     &                    & $1\cdot 10^{-5}$ &  \\
      \texttt{alpha1\_dres} &        &         &      &      &  1120   &  1790   &                    & $1\cdot 10^{-5}$ &  \\
      \texttt{alpha2} &        &         &      &      &     &     &                    & $1\cdot 10^{-4}$  &  \\
      \texttt{alpha3} &        &         &      &      &     &     &                    & $5\cdot 10^{-4}$  & \\
      \texttt{alpha3\_dres} &        &         &      &      &  1120   &  1790  &                    & $5\cdot 10^{-4}$ &  \\
      \texttt{alpha4} &        &         &      &      &     &     &                    & $1\cdot 10^{-3}$  &  \\
      \texttt{alpha4\_dres} &        &         &      &      &  1120  &  1790  &                    & $1\cdot 10^{-3}$ &   \\
      \texttt{alpha5} &        &         &      &      &     &     &                    & $2\cdot 10^{-3}$  & \\
      \texttt{alpha6} &        &         &      &      &     &     &                    & $3\cdot 10^{-3}$  &  \\
      \texttt{alpha6\_dres} &        &         &      &      &  1120  &  1790  &                    & $3\cdot 10^{-3}$ &   \\
     [0.2em]\hline
    \end{tabular}
    \tablefoot{
      Blank spaces assume an identical parameter as the fiducial model
      \texttt{fid}. The simulation labels starting with \texttt{hr} denote
      models with a variation in aspect ratio. The models starting with
      \texttt{p} include a change in the mass of planet~1 or 2 depending on
      the following digit. A suffix of \texttt{fb} describes models with dust
      feedback activated. All models use a radially varying alpha viscosity
      model with $\alpha = 2\cdot 10^{-4}$ except the simulations denoted by
      \texttt{alpha1} to \texttt{alpha6} where a radially constant value of
      $\alpha$ was chosen.
      }
  \end{center}
\end{table*}

\begin{table*}[ht]
  \begin{center}
    \caption{\label{tab:density_normalization}
    Optical depth fitting results for ring~1.}
    \begin{tabular}{lcccccc}
      \hline\hline                       \\[-0.6em]
      Constant & $\Sigma_{\mathrm{g}, 0}\,(r = 48\,\mathrm{au})$ [$\frac{g}{cm^2}$] & $\Sigma_{\mathrm{g}, 0}\,(r = 1\,\mathrm{au})$ [$\frac{g}{cm^2}$] & $a_\mathrm{min}$ [mm] & $a_\mathrm{max}$ [mm] & $\kappa_\mathrm{min}$ [$\frac{cm^2}{g}$] & $\kappa_\mathrm{max}$ [$\frac{cm^2}{g}$] \\[0.2em]
      \hline                             \\[-0.6em]
      Grain size & 37.4 & 827.7 & 0.191 & 19.1 & 1.32 & 0.11 \\
      d/g ratio  & 1.31 & 28.91  & 0.007 & 0.67 & 0.40 & 3.30 \\[0.2em]
      \hline
    \end{tabular}
    \tablefoot{
      Results for both the high and low mass model as derived from the optical depth fitting of ring~1 in Sec.~\ref{sec:rings}.
      Both the initial gas surface densities $\Sigma_\mathrm{g, 0}$ at 1 and 48~au are listed. 
      Depending on the gas density a different dust grain size distribution with the minimum and maximum size of $a_\mathrm{min}$ and $a_\mathrm{max}$ was chosen. 
      Their respective opacity values are listed under $\kappa_\mathrm{min}$ and $\kappa_\mathrm{max}$.
    }
  \end{center}
\end{table*}

\begin{table*}[ht]
  \begin{center}
    \caption{Gaussian fit results of the simulated dust rings for all Stokes numbers, $\mathrm{St}$.}
    \label{tab:ring_fits}
    \begin{tabular}{cccccccc}
      \hline\hline                       \\[-0.6em]
      a [mm] & St & $w_\mathrm{ring1}$ [au] & $w_\mathrm{ring2}$ [au] & $r_\mathrm{ring1}$ [au] & $r_\mathrm{ring2}$ [au] & $M_\mathrm{ring1}$ [$M_\mathrm{earth}$] & $M_\mathrm{ring2}$ [$M_\mathrm{earth}$] \\[0.2em]
      \hline                             \\[-0.6em]
  0.2 & $1.3 \cdot 10^{-3}$ & $4.09 \pm 0.09$ & $10.5 \pm 0.17$ & $62.26 \pm 0.09$ & $100.18 \pm 0.16$ &  2.1     &  1.9                         \\
  0.4 & $2.6 \cdot 10^{-3}$ & $3.13 \pm 0.08$ & $7.10 \pm 0.09$ & $62.05 \pm 0.08$ & $99.90 \pm 0.09$ &  2.9     &  2.9                         \\
  0.7 & $5.0 \cdot 10^{-3}$ & $2.27 \pm 0.07$ & $4.80 \pm 0.04$ & $61.90 \pm 0.07$ & $99.25 \pm 0.04$ &  4.0     &  4.7                        \\
  1.4 & $9.6 \cdot 10^{-3}$ & $1.68 \pm 0.05$ & $3.34 \pm 0.01$ & $61.86 \pm 0.05$ & $98.87 \pm 0.01$ &  5.5     &  8.0                       \\
  2.6 & $1.9 \cdot 10^{-2}$ & $1.28 \pm 0.04$ & $2.37 \pm 0.01$ & $61.89 \pm 0.04$ & $98.69 \pm 0.01$ &  7.9    &  12.3                       \\
  5.1 & $3.6 \cdot 10^{-2}$ & $1.00 \pm 0.03$ & $1.76 \pm 0.01$ & $61.92 \pm 0.03$ & $98.67 \pm 0.01$ &  11.5    &  17.0                       \\
  9.9 & $6.9 \cdot 10^{-2}$ & $0.80 \pm 0.03$ & $1.33 \pm 0.01$ & $61.93 \pm 0.03$ & $98.67 \pm 0.01$ &  17.3    &  22.2                      \\
  19.1 & $1.3 \cdot 10^{-1}$ & $0.64 \pm 0.02$ & $1.01 \pm 0.01$ & $61.94 \pm 0.02$ & $98.66 \pm 0.01$ &  28.0    &  26.9                     \\
  sum (high mass model)   &                 &                 &                  &         &          &  79.3   &  95.9 \\
  sum (low mass model)   &                 &                 &                  &       &            &  11.4   &  13.8 \\[0.2em]
      \hline
    \end{tabular}
    \tablefoot{
      Ring widths are denoted by $w_\mathrm{ring1,2}$ while the ring position is labeled $r_\mathrm{ring1,2}$. 
      The dust size individual mass trapped in the rings and the total mass are shown in the last two columns.
    }
  \end{center}
\end{table*}

\subsection{Radiative transfer model and post processing} \label{sec:radtrans}
In order to compare the results of the hydrodynamical simulations with the
observational data synthetic images are produced with RADMC-3D
\citep{Dullemond2012} and the CASA package \citep{mcmullin_casa_2007}. The
results of the hydrodynamical simulations have to be extended to three
dimensional dust density models which then serve as input for the radiative
transfer calculations with RADMC-3D.\\ Using the given Stokes numbers of the
hydro model, the respective dust sizes are computed via
Eq.~\ref{eq:stokes_number}. The number density size distribution of the dust
grains follows the MRN distribution $n(a) \propto a^{-3.5}$
\citep{mathis_1977} where a is the grain size. \\ Dust settling towards the
mid-plane is considered following the diffusion model of
\cite{dubrulle_dustsettling_1995} 
\begin{equation} \label{eq:dust_settling}
H_\mathrm{d} = \sqrt{\frac{\alpha}{\alpha + \mathrm{St}}} H 
\end{equation}
Here, we assume a Schmidt number on the order of unity. \revised{
  The grid resolution for the radiative transfer is identical to the hydrodynamical mesh.
  In polar direction the grid is expanded by 32 cells which are equally spaced up to $z_\mathrm{lim} = \pi/2
\pm 0.3$.
}  The vertical disk density profile is assumed to be isothermal and
the conversion from the surface density to the local volume density is
calculated as follows: 
\begin{equation} 
 \rho_\mathrm{cell} =
\frac{\Sigma}{\sqrt{2\pi} H_\mathrm{d}} \cdot \mathrm{erf}^{-1} \left(
\frac{z_\mathrm{lim}}{\sqrt{2} H_\mathrm{d}} \right) \cdot \frac{\pi}{2}
H_\mathrm{d} \frac{\ \left[ \mathrm{erf}\left( \frac{z_+}{\sqrt{2}
H_\mathrm{d}} \right) - \mathrm{erf}\left( \frac{z_-}{\sqrt{2} H_\mathrm{d}}
\right) \right]}{z_+ - z_-} . 
\end{equation} 
The error function term is a
correction for the limited domain extend in the vertical direction that would
otherwise lead to an underestimation of the total dust mass. Similarly, the
second correction term accounts for the finite vertical resolution,
especially important for thin dust layers with strong settling towards the
mid-plane. The coordinates $z_+$ and $z_-$ denote the cell interface
locations in polar direction along the numerical grid.\\
For each grain size
bin and a wavelength of 1.3~mm the corresponding dust opacities were taken
from the \texttt{dsharp\_opac} package which provides the opacities presented
in \cite{birnstiel_dsharp_2018}. These opacities are based on a mixture of
water ice, silicate, troilite and refractory organic material.
\revised{
The grains are assumed to be spherically shaped and to have no porosity. In
the RADMC-3D model, the central star is assumed to have a mass of 1.9 solar
masses with an effective temperature of 9333 K which results in a luminosity
of $2.62 \cdot 10^{34} \mathrm{erg} / \mathrm{s}$. The system is assumed to
be at a distance of 101 pc. For the dust temperature calculation a number of
$n_\mathrm{phot} = 10^8$ photon packages and for the image reconstruction
$n_\mathrm{phot\_scat}=10^7$ photon packages were used. The thermal
Monte-Carlo method is based on the recipe of \cite{bjorkman_wood_2001}. We
include isotropic scattering in the radiative transfer calculation. A
comparison between the prescribed gas temperatures and the computed dust
temperatures is given in Appendix \ref{sec:appendix_dust_temperatures}.} \\
For simulating the
detectability of the various features present in the model we use the task
\texttt{simalma} from the CASA-5.6.1 software. A combination of the antenna
configurations \texttt{alma.cycle4.8} and \texttt{alma.cycle4.5} was chosen.
The simulated observation time for configuration 8 is 2 hours while the more
compact configuration is integrated over a reduced time with a factor of 0.22
corresponding 0.44 hours.\\ We employed the same cleaning procedure as made
available in \cite{Isella2018} to reduce the artificial features from the
incomplete uv-coverage and to allow a comparison to the observation. The
procedure involves the CASA task \texttt{tclean} with a robust parameter of
-0.5 and manual masking of the disk geometry. Consistent with the radiative
transfer model, the observed wavelength is simulated to be at $1.3 \,
\mathrm{mm}$, corresponding to ALMA band 6. \\

\section{Results} \label{sec:results}
In the following parts the outcome of the simulation runs listed in 
Table~\ref{tab:simulations} will be presented and analyzed.
First, the variety of substructures emerging from the interaction of gas,
dust and the three planets will be described. Afterwards parameter dependence
and observability will be addressed.

\subsection{Dust substructure overview}
\begin{figure*}[ht] 
  \centering
  \includegraphics[width=\linewidth]{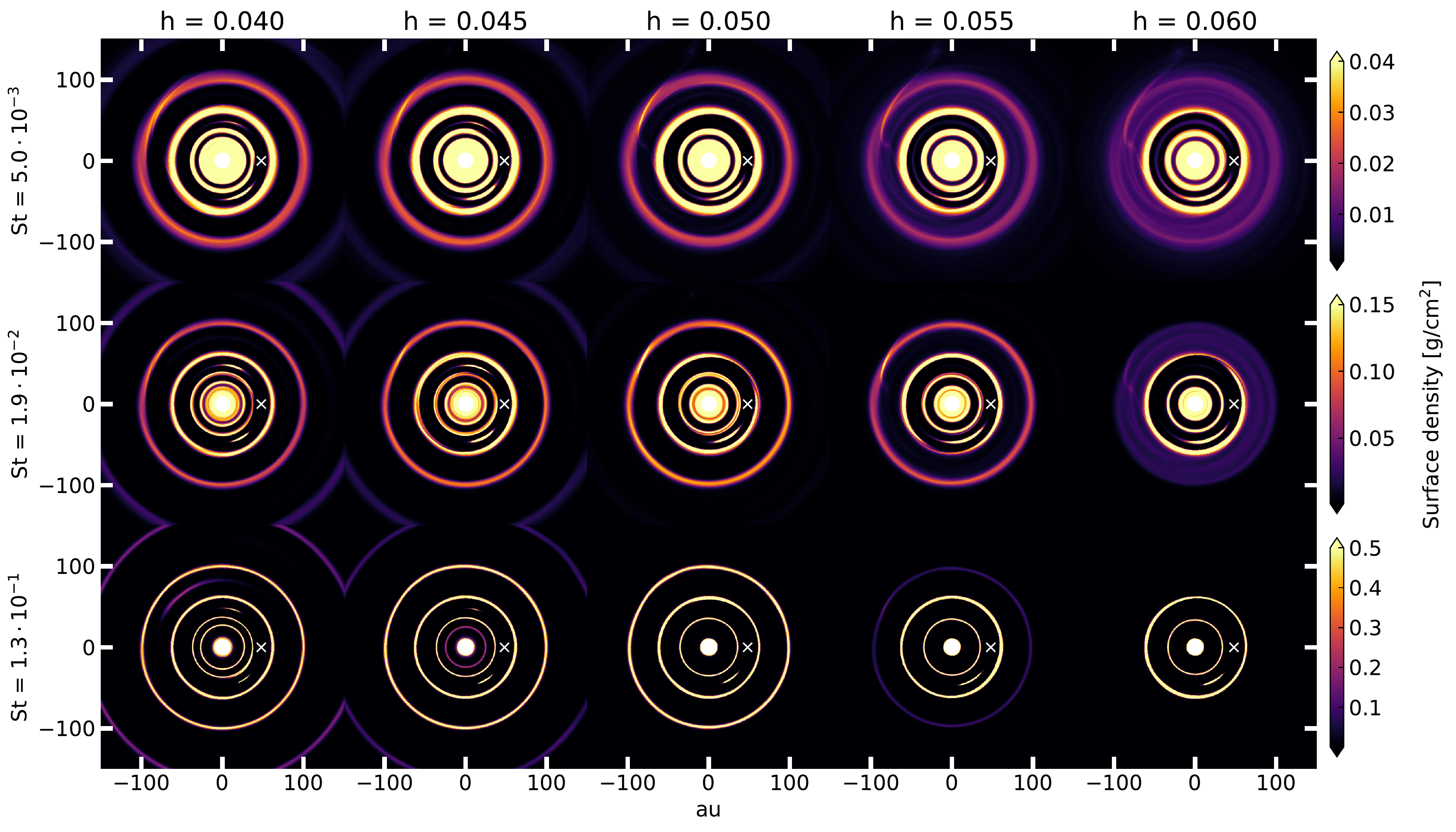}
  \caption{Shown are dust surface density maps for a subset of three fluids
  with varying values of the aspect ratio $h = H/r$ at 500 orbits at 48~au.
  \revthird{Crescent-shaped asymmetries} are visible for all aspect ratios. 
  \revsecond{
  Large values of $h$ weaken dust accumulation in the co-orbital regions of the planets.
  The white crosses mark the position of planet~1.}
  }
  \label{fig:dust_hr}
\end{figure*}

\begin{figure*}[ht] 
  \centering
  \includegraphics[width=\linewidth]{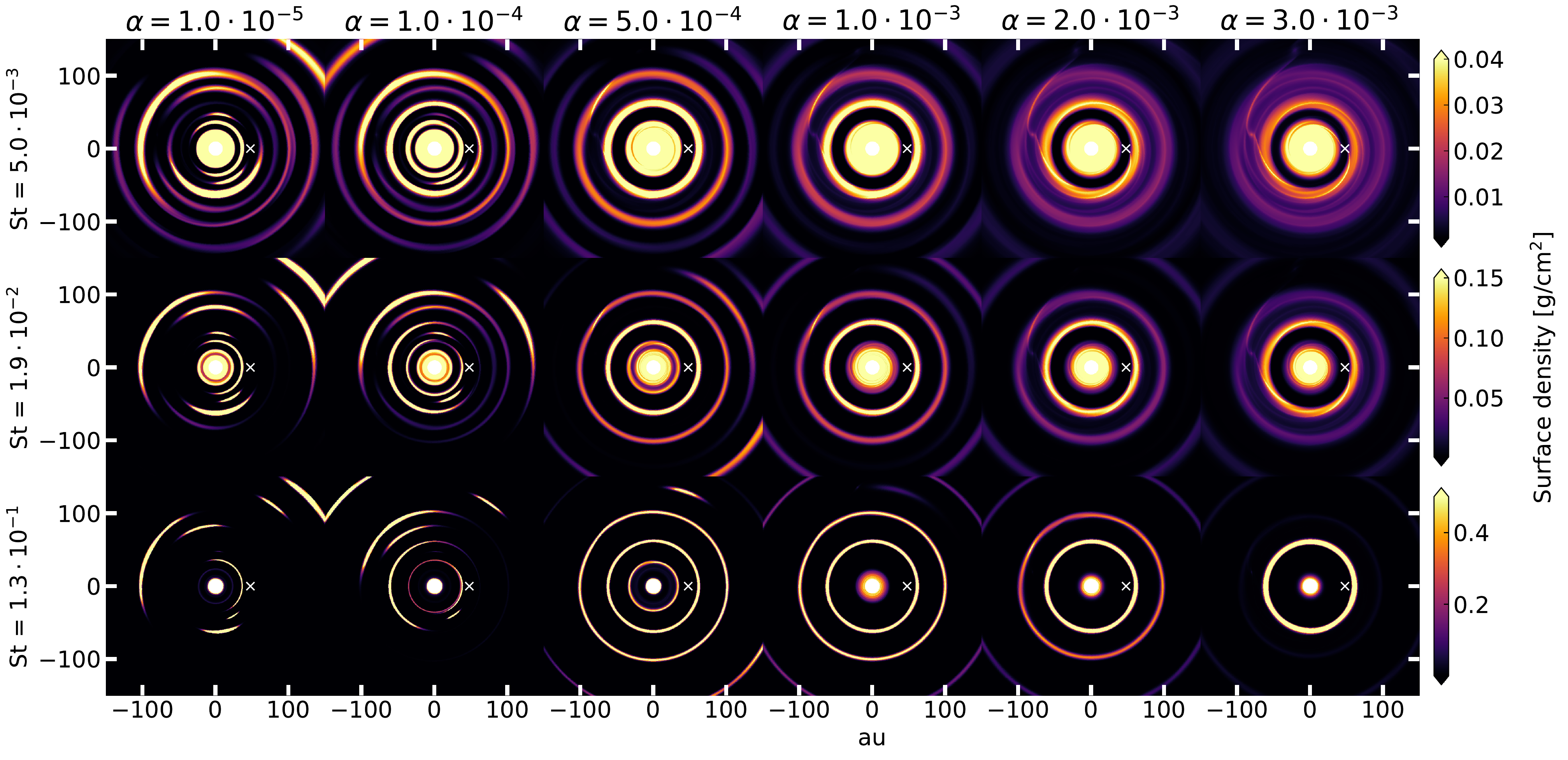}
  \caption{Dust surface density maps for different values of a radially
  constant $\alpha$ viscosity after 500 orbits at 48~au. For $\alpha \leq
  10^{-4}$ strong asymmetries are present. ring~2 weakens or vanishes for
  large viscosities.
  \revsecond{The white crosses mark the position of planet~1.}
  }
  \label{fig:dust_alpha}
\end{figure*}

A variety of dust substructures emerges from the planet-disk system during
its dynamical evolution. Fig.~\ref{fig:dust_hr} and Fig.~\ref{fig:dust_alpha}
show the dust surface density structure for a selected parameter space of
aspect ratio and the turbulent viscosity, characterized by $\alpha$.
\revsecond{In the following analysis of the \revthird{crescent-shaped
asymmetries} simulation snapshots after 500 orbits at 48 au are compared with
each other since their evolution is comparable for all resolutions.}
Most
prominently multiple rings form in most cases. As expected, fluids with
larger Stokes numbers $\mathrm{St}$ and thus larger grain sizes exhibit
thinner rings and more concentrated substructures. Especially,
nonaxisymmetric features can be seen mostly for $a \geq 2.6\,\mathrm{mm}$ or
$\mathrm{St} > 10^{-2}$. For smaller dust sizes the dust is better coupled to
the gas and resembles its structure more closely. \\ 
\revsecond{Additionally, if \revthird{a crescent-shaped
asymmetry} is present, it is situated in the gap caused by the most inner
planet at 48~au for the majority of the parameter space. Further asymmetries
appear if the $\alpha$-viscosity is radially constant with values of $\alpha
\leq 1\cdot 10^{-3}$ or for low values of the aspect ratio $h$. }
Dust is preferably trapped in the Lagrange point L5. In
Fig.~\ref{fig:dust_hr} rings and asymmetries become weaker with increasing
aspect ratio. 
\revised{
  A crescent-shaped feature at the L5 position is present for all aspect
  ratios while a second similar asymmetry at the L4 point appears for values of H / r
  < 0.05. The crescent weakens for smaller grain sizes.
}
Also the second prominent ring beyond planet~2 at 83~au
weakens clearly for larger aspect ratios. In the combination of the largest
grain sizes and $H / r$ the ring completely vanishes. \\ To highlight the
importance of the turbulent $\alpha$ viscosity parameter a subset of results
with radially constant values of $\alpha$ are shown in
Fig.~\ref{fig:dust_alpha}. Not surprisingly, larger values of alpha generally
lead to a more diffuse and symmetric distribution of dust. Below $\alpha = 5
\cdot 10^{-4}$ no concentric rings form due to vortices in the gas.
Crescent-shaped features in both Lagrange points of the innermost planet are
visible
in the very low viscosity case of $\alpha = 10^{-5}$. A sufficiently large
viscosity on the other hand also leads to the disappearance of the second
ring in the limit of larger grains and Stokes numbers, similar to the large
aspect ratio in Fig.~\ref{fig:dust_hr}. The collection of these results also
stresses the issue with a radially constant $\alpha$ viscosity with respect
to the observed HD~163296 system. In order to reproduce a nonaxisymmetric
feature in the vicinity of the inner planet and a smooth ring-shaped outer
structure, a radially increasing value of $\alpha$ would be the natural
choice. This is also consistent with an embedded dead zone at the inner part
of the disk and a more active outer disk region with a higher degree of
ionization \citep{miranda_2016, pinilla_2016}. We thus chose a radially
increasing $\alpha$ viscosity for all remaining simulation runs. \\ Models
with a variation in the dust cutoff radius only show little changes in the
resulting dust structures. Only this subset of the possible parameter space
already exposes the degeneracy of the emerging substructures with respect to
the chosen disk models.

\subsection{Rings} \label{sec:rings}

\begin{figure}[ht] 
  \centering
  \includegraphics[width=\linewidth]{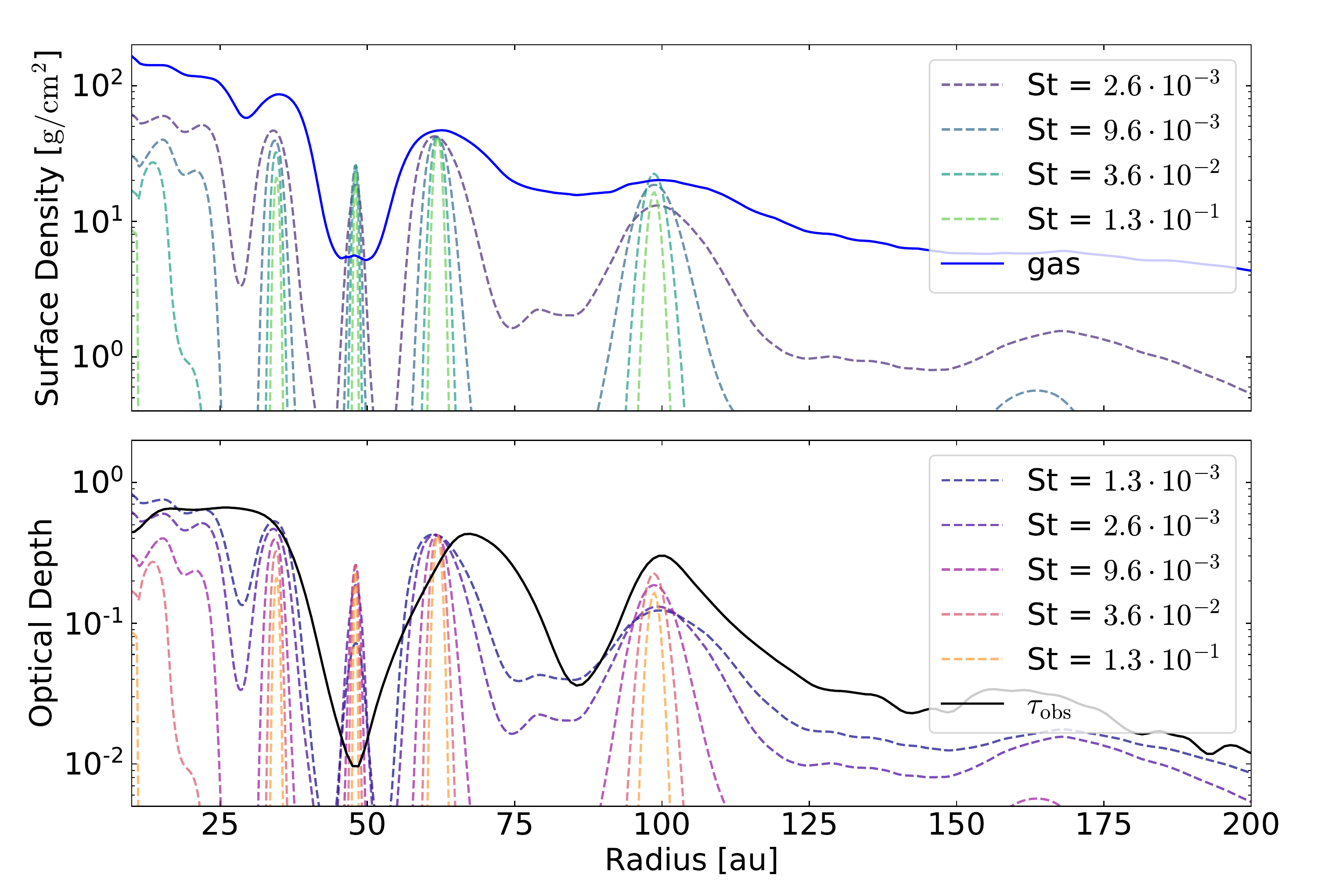}
  \caption{\textit{Upper panel:} azimuthally averaged dust and gas surface
  densities of model \texttt{fid\_dres} after 1000 orbits at 48~au. Dust density
  profiles are normalized to the gas peak value at the location of ring~1. \\
  \textit{Lower panel:} azimuthally averaged optical depths of model
  \texttt{mid} after 1000 orbits at 48~au compared to the observed optical
  depth $\tau_\mathrm{obs}$. Simulated optical depths are normalized to
  $\tau_\mathrm{obs}$ at the location of ring~1.
  }
  \label{fig:surface_dens_tau}
\end{figure}

\begin{figure}[ht] 
  \centering
  \includegraphics[width=\linewidth]{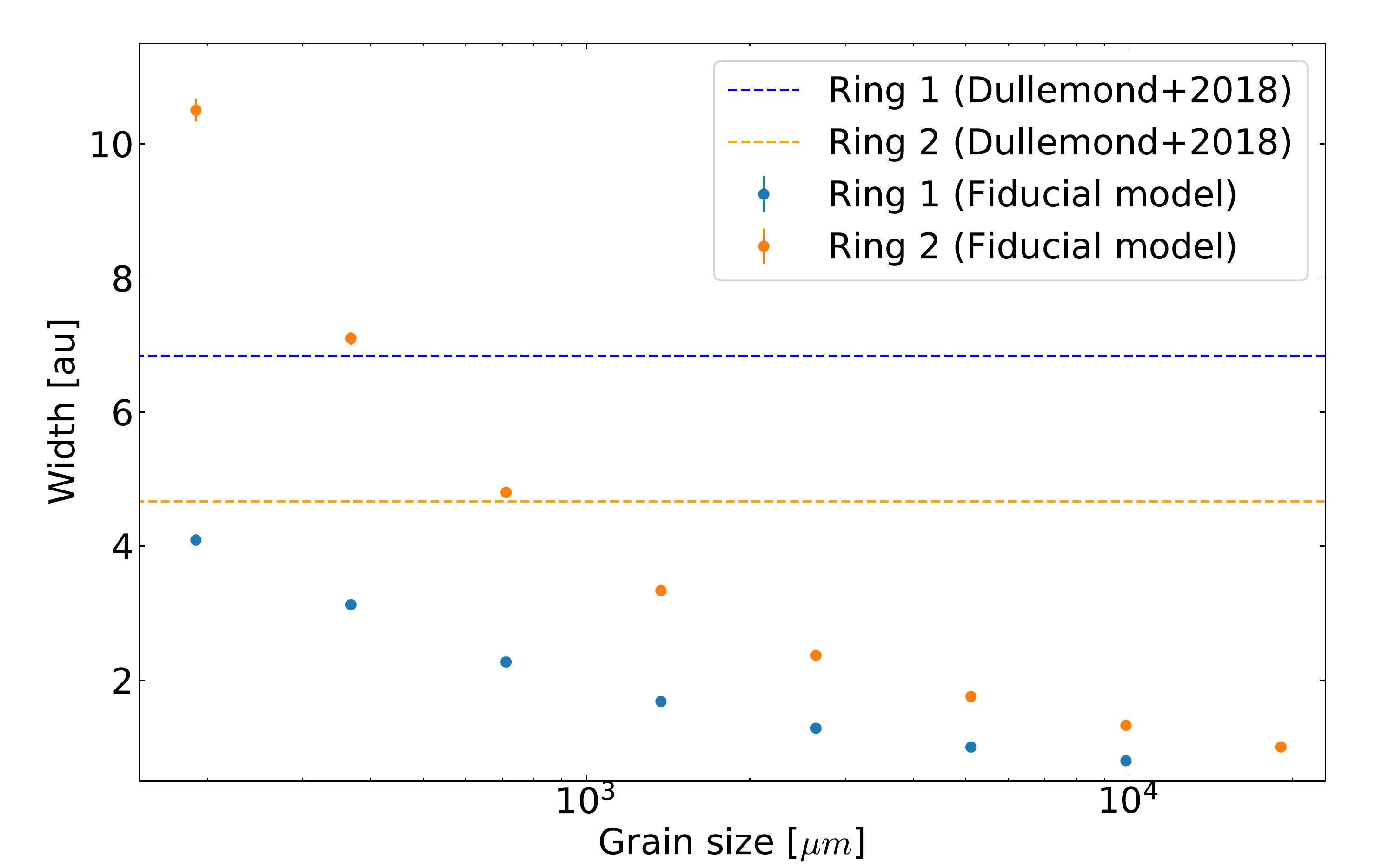}
  \caption{Ring widths of model \texttt{fid\_dres} after 1000 orbits at
  48~au. Simulated values are compared to the inferred ring widths of
  \cite{Dullemond2018}, displayed as the horizontal dashed lines.}
  \label{fig:ring_width}
\end{figure}

\begin{figure*}[ht] 
  \centering
  \includegraphics[width=\linewidth]{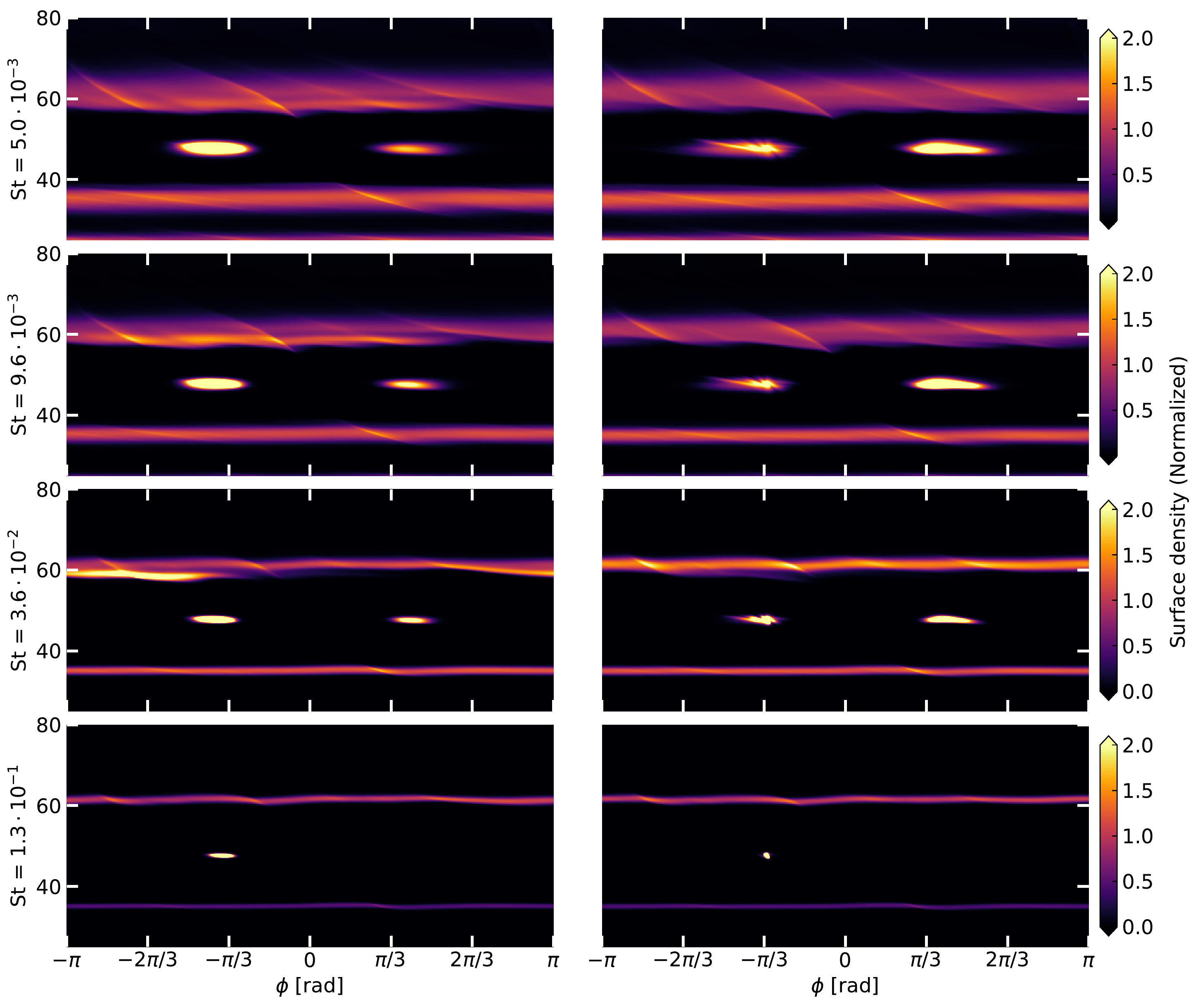}
  \caption{
  \revised{
  Dust surface density maps in polar coordinates for model
  \texttt{fid\_dres} without dust feedback (left hand side) and
  \texttt{p1m6fb\_dres} with dust feedback after 500 orbits at 48~au for four
  different \revthird{initial Stokes numbers} and dust sizes. A dust agglomeration around the location of the
  trailing Lagrange point L5 is present for all Stokes numbers. 
  Unlike the nonfeedback case the right
  panels show that dust is trapped more efficiently in the leading Lagrange point L4 for
  smaller Stokes numbers. 
  \revthird{
  For $\mathrm{St} \lesssim 3.6 \cdot 10^{-2}$ the L4 feature is more pronounced than the dust over
  density around L5. Dust feedback leads to an instability at the L5 point and fragments the asymmetric feature.
  Dust densities are normalized to the peak value of ring~1.
  }
  }}
  \label{fig:asymmetry_polar}
\end{figure*}


\begin{figure}[ht] 
  \centering
  \includegraphics[width=\linewidth]{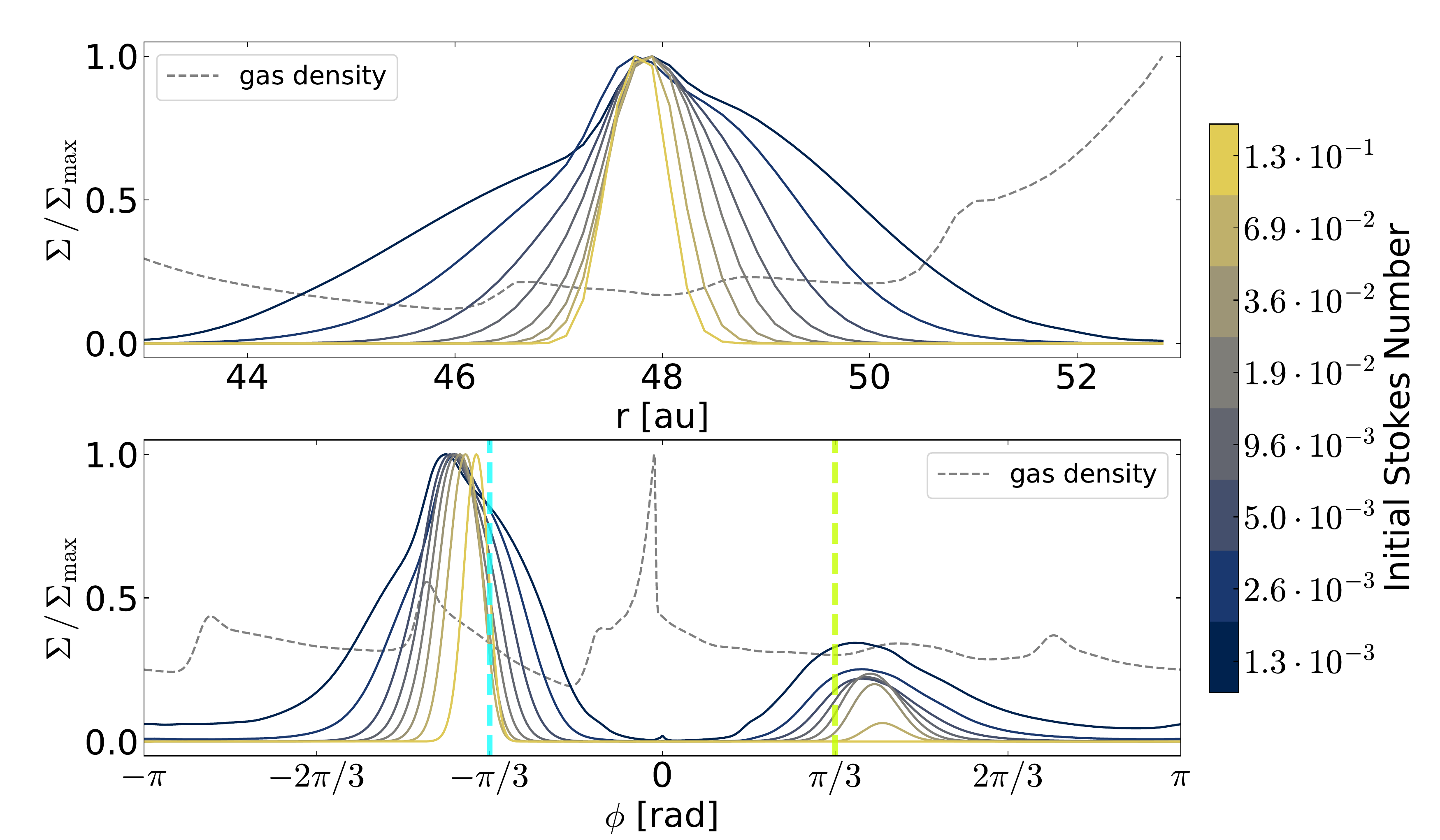}
  \caption{Radial and azimuthal cut through the maximum of the \revthird{crescent-shaped asymmetry}
  around L5 of the fiducial model \texttt{fid\_dres} after 500 orbits at 48~au. The
  color map indicates the different \revthird{initial Stokes numbers} of the dust
  fluids. 
  \revsecond{The dashed lines represent the gas density.
  The surface densities are normalized to their respective maximum value in the co-orbital region of planet~1.}
  The light blue and light green vertical lines indicate the Lagrange points L5
  and L4 respectively.}
  \label{fig:asymmetry_width}
\end{figure}

\begin{figure}[ht] 
  \centering
  \includegraphics[width=\linewidth]{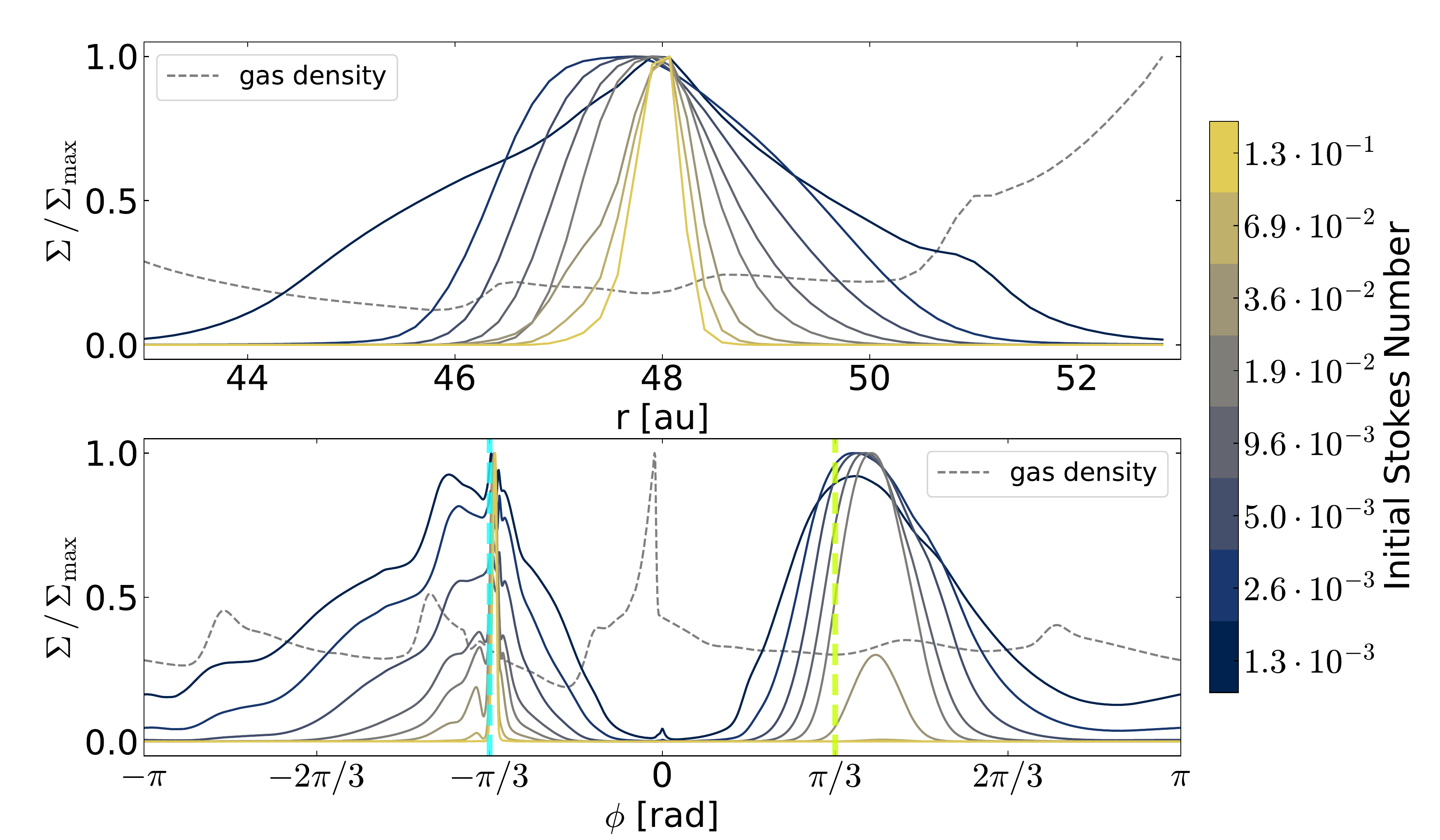}
  \caption{Radial and azimuthal cut through the maximum density value in the
  co-orbital region of the model \texttt{p1m6fb\_dres} with dust feedback
  after 500 orbits at 48~au. The light blue and light green vertical lines
  indicate the Lagrange points L5 and L4 respectively. Depending on the
  location of the feature the density cuts are normalized to the peak value
  in the region of L4 or L5. \revthird{Smaller dust with $\mathrm{St} \lesssim 3.6 \cdot 10^{-2}$
  preferably concentrates in L4. }}
  \label{fig:asymmetry_width_fb}
\end{figure}

\begin{figure}[ht] 
  \centering
  \includegraphics[width=\linewidth]{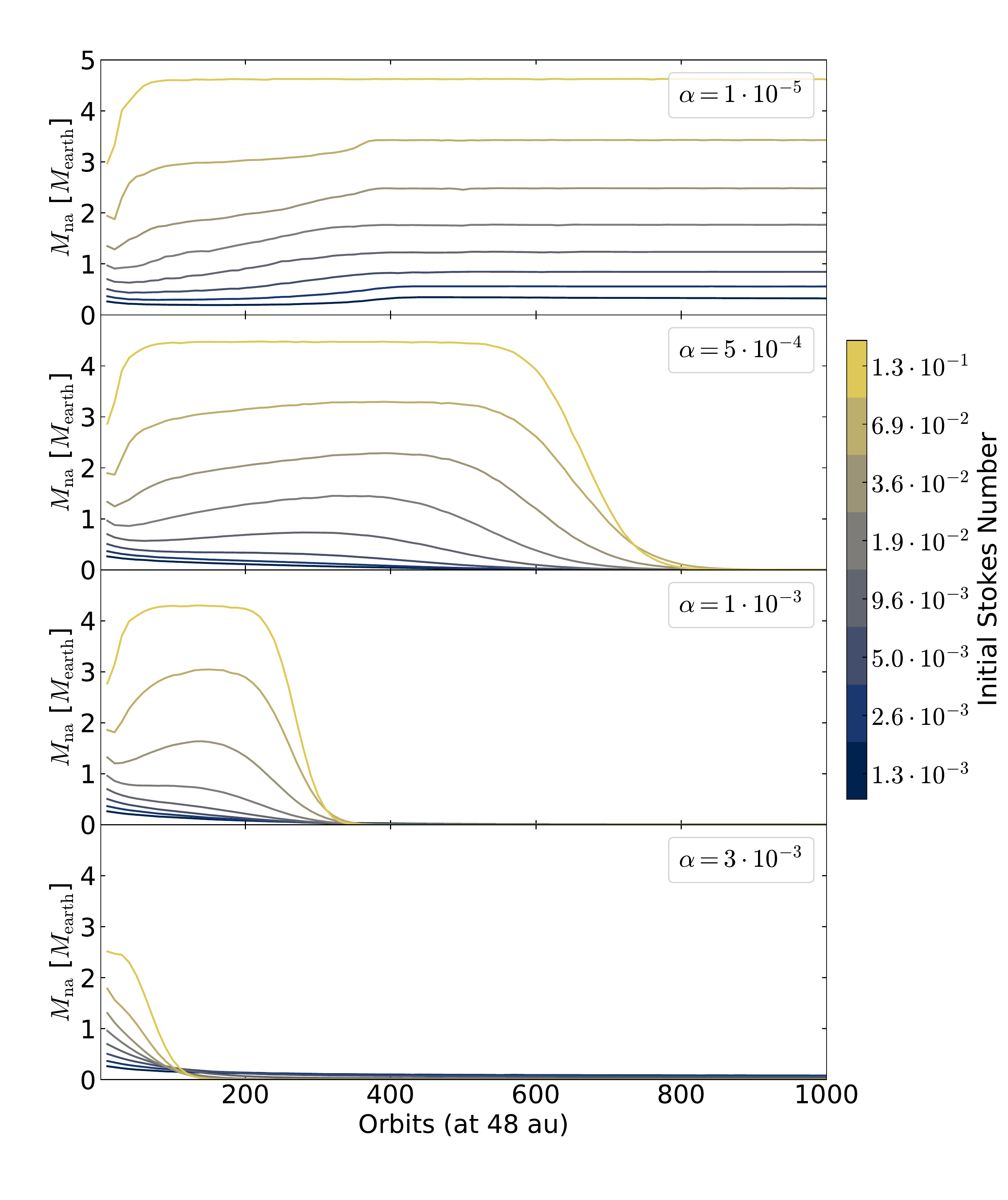}
  \caption{Development of the trapped dust mass in the L5 region for
  different $\alpha$ viscosities over time with a resolution of 14 cells per
  scale height. Dust masses $M_\mathrm{na}$ in
  the nonaxisymmetric feature denoted are integrated down to a cut-off value
  of $10^{-2} \, \Sigma_\mathrm{max}$ and normalized to the high mass model
  stated in Table~\ref{tab:density_normalization}.}
  \label{fig:asymmetry_time_alpha}
\end{figure}

\begin{figure*}[ht] 
  \centering
  \includegraphics[width=\linewidth]{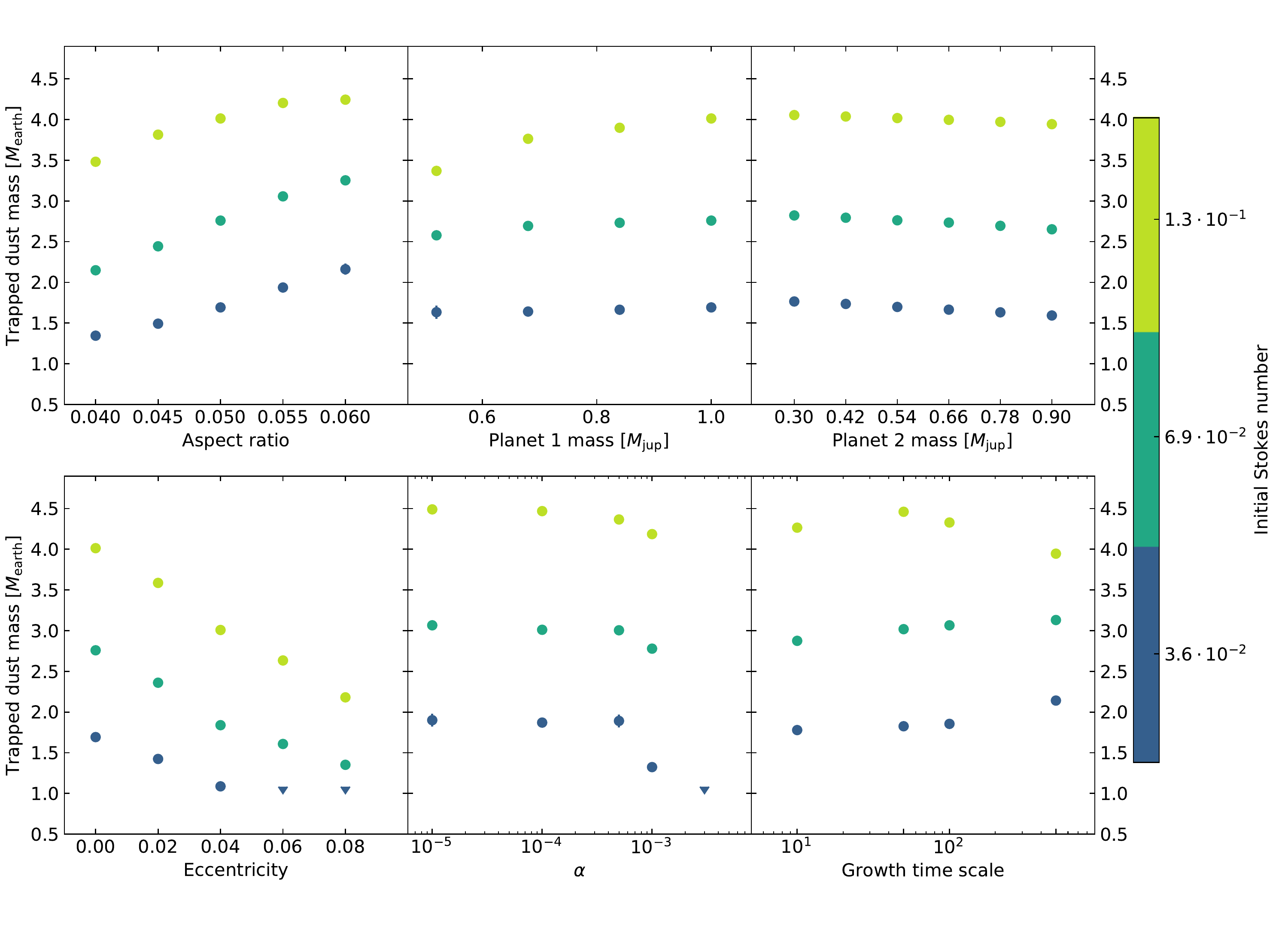}
  \caption{Trapped dust masses in earth masses in the L5 feature for the
  largest three inital Stokes numbers. The mass values are averaged over the
  approximately constant mass time frame starting from $100 \, T_0$. Dust masses are normalized
  according to the high mass model stated in
  Table~\ref{tab:density_normalization}. Missing data points are equivalent
  to a nonexistence of a stable \revthird{crescent-shaped asymmetry} in the co-orbital region around
  48~au.}
  \label{fig:plateau_mass_evolution}
\end{figure*}

\begin{figure*}[ht] 
  \centering
  \includegraphics[width=\linewidth]{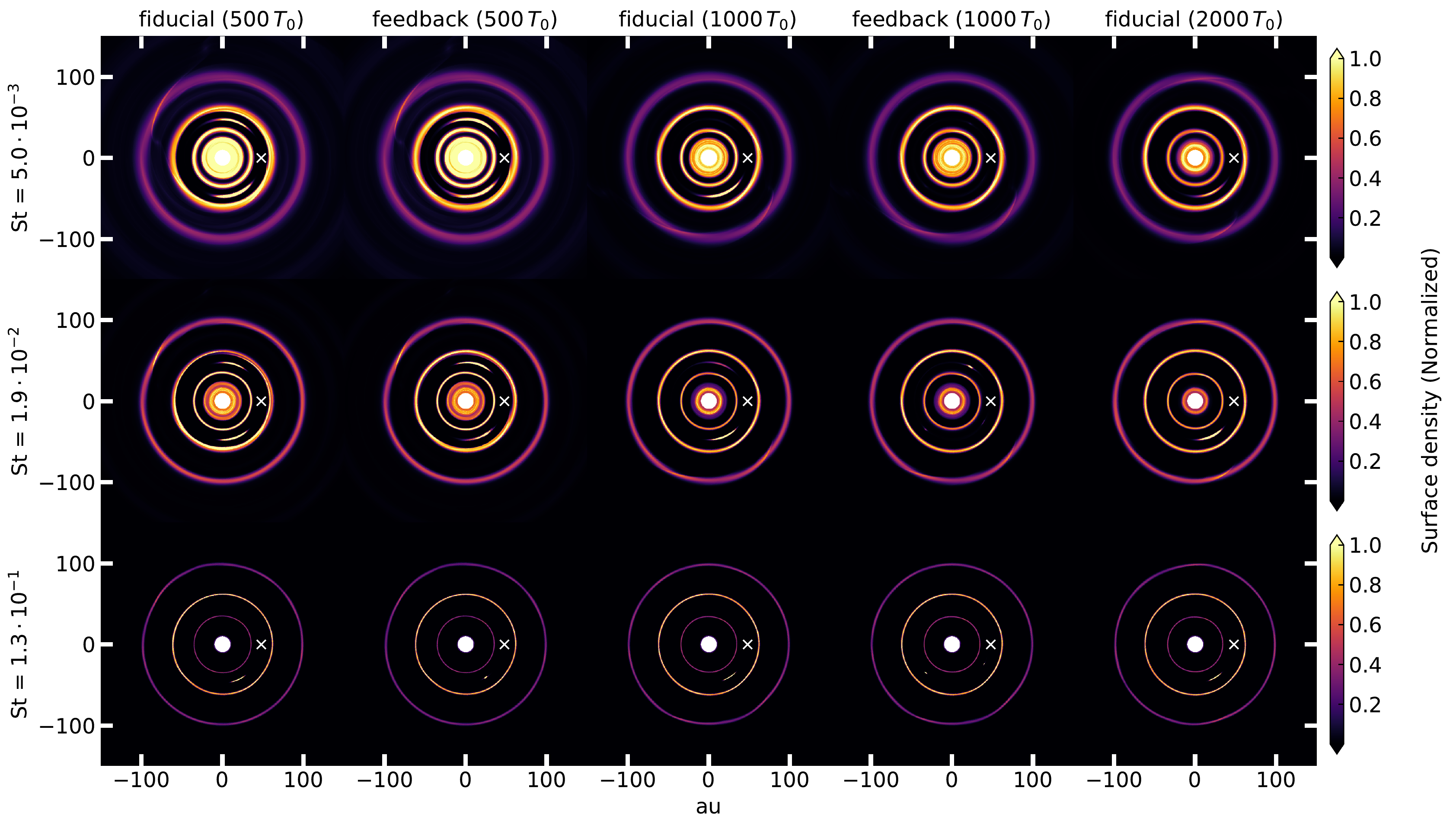}
  \caption{
  \revthird{
    Comparison of the dust surface density map for three different dust sizes
    of the models \texttt{fid\_dres} (fiducial) and \texttt{p1m6fb\_dres}
    (feedback) at different simulation times. The surface densities are normalized to the peak value at ring~1.
  }  
  }
  \label{fig:dust_overview}
\end{figure*}

\begin{figure*}[ht] 
  \centering
  \includegraphics[width=\linewidth]{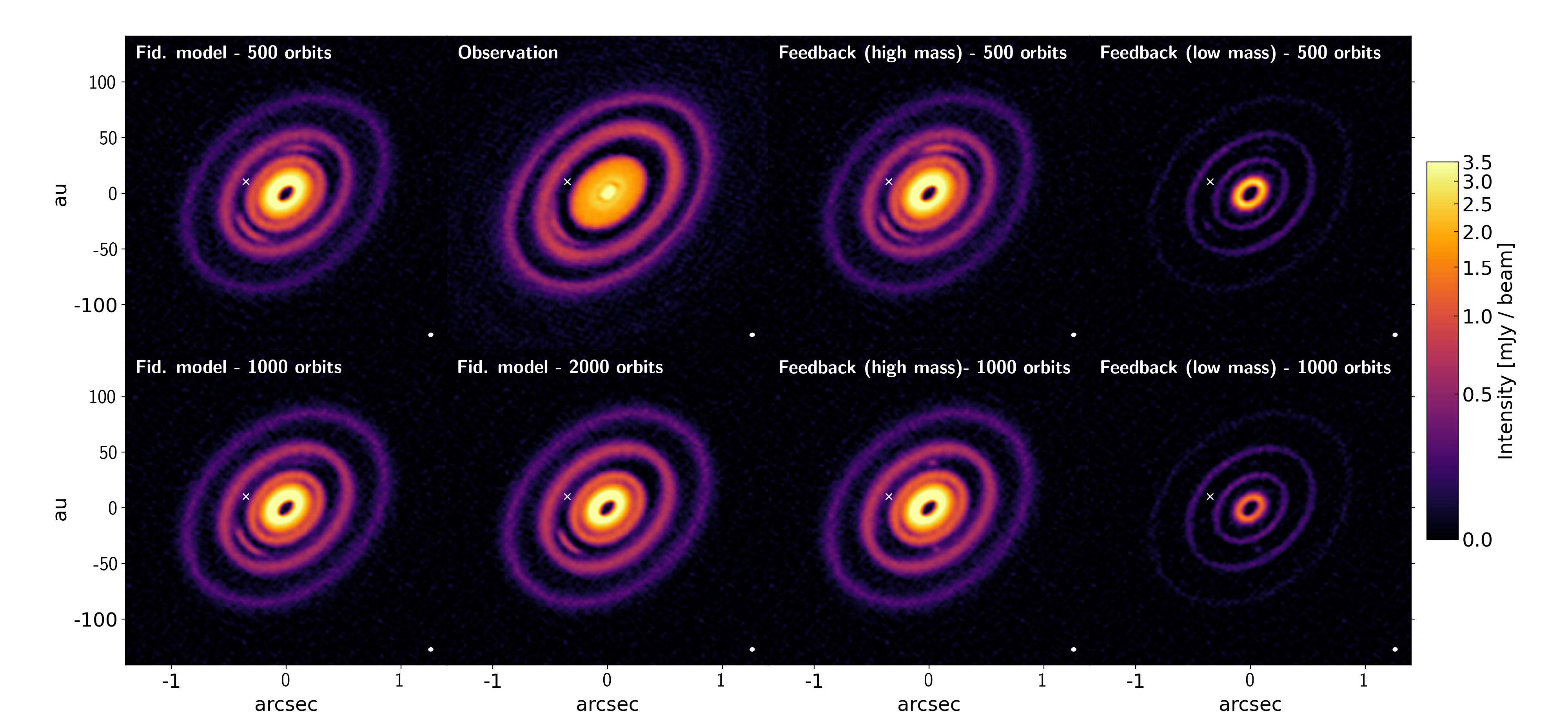}
  \caption{
    \revsecond{Comparison of the synthetic images based on the models
  \texttt{fid\_dres} (first and second column), 
  \texttt{p1m6fb\_dres} (third and fourth column) with the observation taken
  from \cite{Andrews2018, Isella2018}.
  The ellipsis at the lower right of each panel visualizes the synthesized
  beam. The beam size is $49.7 \times 41.4 \,\mathrm{mas}$ for the synthetic
  images. The rms noise reaches $\approx 50 \, \mu\mathrm{as}$. The synthetic
  images are projected with an inclination of $46.7 \degree$ and a position
  angle of $133.33 \degree$.}}
  \label{fig:synth_images}
\end{figure*}
\begin{figure*}[ht] 
  \centering
  \includegraphics[width=\linewidth]{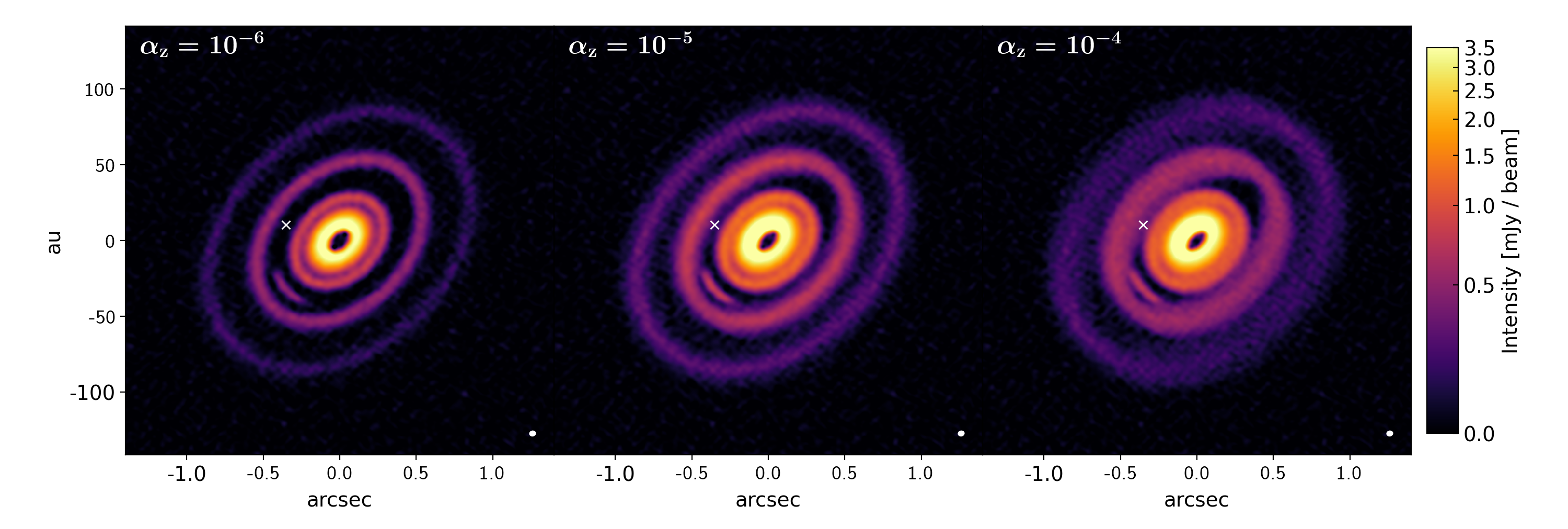}
  \caption{
  \revsecond{  
  Comparison of the synthetic images based on the model
  \texttt{fid\_dres} after 2000 orbits at 48 au with respect to the vertical dust mixing parametrized by $\alpha_\mathrm{z}$.                                    
  }
 }
  \label{fig:synth_images_alpha}
\end{figure*}
Before quantifying the nonaxisymmetric feature in the model, we want to
compare properties of the ring structures with previous works and the
observations. The general procedure is to azimuthally average the dust
density maps after 1000 orbits at 48~au and to invoke Gaussian fitting of the
dust rings, comparable to \cite{Dullemond2018}. 
\revsecond{
Since the ring structure is approximately converged after 900 orbits the following procedure is based on the snapshots at 1000 orbits.
}
\\
Fig.~\ref{fig:surface_dens_tau} shows the results of the  \revthird{high resolution fiducial model
\texttt{fid\_dres} as an example of the radial surface density structure.
The increased resolution was chosen since the lower resolution models
may overestimate the ring width and the trapped dust mass due to numerical
diffusion (appendix~\ref{sec:appendix_res_study}).
} In the
upper panel the dust species is rescaled to gas density peak of ring~1. The
dust rings are clearly thinner than the gaseous envelope and the ring width
decreases with increasing dust grain sizes due to the stronger drift.\\ With
the corresponding opacities $\kappa_\mathrm{i}$ a rough estimate of the
resulting optical depth can be computed by $\tau_\mathrm{sim} =
\kappa_\mathrm{i} \Sigma_\mathrm{i}$, where i denotes the dust species index.
The results of this estimate are displayed in the lower panel of
Fig.~\ref{fig:surface_dens_tau}. All optical depths are rescaled to the
values at the position of ring~1 from the profile $\tau_\mathrm{obs}$ derived
in \cite{Huang2018a}. The profile provided in their work excludes
contributions from the prominent nonaxisymmetric structures. Several
properties become apparent:
\begin{itemize}
    \item Ring~1 is wider than in the simulated profile. 
    \item The peak location of ring~1 is located further outward with respect to the simulated one.
    \item The peak value of the optical depth of ring~2 from smaller grains and Stokes numbers is lower in the simulations compared to the estimated value from the observation.
\end{itemize}
A partial explanation of these differences could be that the dust-to-gas
ratio could become larger than in the models so that dust feedback shifts the
ring further outward and spreads the ring as shown in \cite{weber_2018} and
\cite{kanagawa_2018}. Here, the dust-to-gas ratio is not sufficient to cause
a significant effect in the models including dust feedback.\\ A conclusion of
\cite{Dullemond2018} was that the optical depth observed in the DSHARP survey
were remarkably close to unity and that the rings were optically thin. Later
\cite{zhu2019_dust_scattering} argued that dust scattering could account for
this phenomenon and that the actual optical depth could be larger. In the
case of HD~163296 the mass hidden in ring~1 could be thus larger than
expected.\\
For ring fitting we use a Gaussian:
\begin{equation}
    \Sigma_\mathrm{fit}(r) = A \, \mathrm{exp} \left( - \frac{(r - r_0)^2}{2w} \right)
\end{equation}
with the peak value $A$, the ring location $r_0$ and the ring width $w$.
In Fig.~\ref{fig:ring_width} the fitted ring widths from model \texttt{fid}
are plotted and compared to the observed values in \cite{Dullemond2018}.
Grain sizes of the high mass model are used. The width of ring~2 is matched
close to a grain size of $1\,\mathrm{mm}$. On the other hand, the width of ring~1 is not reached with the
parameters chosen in our models. It should be noted that the gas ring width
is about 8~au, just slightly larger than the observed value of about 7~au.
Smaller Stokes numbers could in principle reproduce these findings. \\ The
equivalent model \texttt{p1m6fb\_dres} including dust feedback shows no significant
differences.

\subsection{Surface density estimation}
With ring~1 being the most prominent substructure in the observed system, its
estimated lower bound for the surface density would be a reasonable choice
for rescaling the simulated dust density maps. Furthermore, the
crescent-shaped feature of interest is located closely to ring~1. In order to
estimate a minimum mass of trapped dust in this feature an appropriate
normalization of the density with respect to ring~1 would be a natural
choice.\\ There are two possible methods in achieving a simple normalization.
First, we rescale the sum of all azimuthally averaged dust densities so that
the combined optical depth at the location of ring~1 equals the observed
value. To maintain the validity of the dynamics of the system, the Stokes
number corresponding to the grain size of a fluid has to be unmodified by
this process. The immediate consequence is then a change in the dust-to-gas
ratio if the densities are rescaled, since a change in the gas surface
density with a constant grain size would modify the Stokes number (see
Eq.~\ref{eq:stokes_number}). With a change in the dust-to-gas ratio the dust
dynamics only remain comparable if no dust feedback is considered.\\ The
second choice would be to maintain an initial dust-to-gas ratio of 0.01 and
to change the dust grain size and the gas surface density. A modification of
the grain size affects in turn the dust opacities and thus the optical depth.
Consequently, the process of generating $\tau_\mathrm{sim}$, rescaling it to
$\tau_\mathrm{obs}$ and inferring the corrected dust surface densities has to
be iterated until convergence is achieved. Keeping the dust-to-gas ratio
constant is important for the model runs with dust feedback enabled. \\
Table~\ref{tab:density_normalization} provides relevant results from these
two approaches which will be denoted by the high and low mass model in the
following parts. With the unchanged grain sizes the dust-to-gas ratio
diminishes to $\approx 2.4 \cdot 10^{-3}$. For the iterative approach with
the dust-to-gas ratio unchanged, the grain size distribution shifts towards
smaller grains with a maximum size \revthird{ $a_\mathrm{max} \approx 0.67\, \mathrm{mm}$
and a minimum size $a_\mathrm{min} = 0.007 \,\mathrm{mm}$.} \\ Results of the
Gaussian fits onto the dust rings of the fiducial model are listed in
Table~\ref{tab:ring_fits} for all simulated dust species. The inferred ring
widths are decreasing with increasing grain sizes and Stokes numbers. The
peak maxima shift towards the star for larger grain sizes since the dust
drift becomes more dominant in this regime. The total dust mass of all
species is computed for both the low and high mass model.
\cite{Dullemond2018} found masses of $56 \, M_\mathrm{earth}$ and $43.6\,
M_\mathrm{earth}$ for ring~1 and ring~2 respectively while assuming 1~mm
grains with equal opacities values as used in the model presented here. The
results of both the low and high mass model encompass the values of
\cite{Dullemond2018}.\\ The high mass model will be the preferred choice in
the following diagrams and analysis since the opacity is dominated by smaller
grain sizes compared to the low mass model. The choice is motivated by the
broad ring structures apparent in the observations.

\subsection{Secondary planet mass} \label{sec:secondary_planet}
In the models \texttt{p2m1} - \texttt{p2m6} the mass of the secondary planet
at 83~au is varied in order to verify its impact on the ring structure. The
results of \cite{Teague2018} indicate a planet mass of $1 \,M_\mathrm{jup}$
within an error margin of $50 \%$. \\ 
\revised{ In our model runs we chose a mass of
$0.55 \,M_\mathrm{jup}$ for planet~2 since a larger mass causes a stronger
dissipation of the \revthird{crescent-shaped asymmetry}. Lower masses
significantly decreased the dust content in
ring~2 and thus the fiducial planet mass value was chosen as the sweet spot
between a strong ring contrast and an maximized asymmetric dust accumulation. Further
details are given in appendix~\ref{sec:appendix_planet2}.}

\subsection{Asymmetries} \label{sec:asymmetries}

Of particular interest is the crescent-shaped asymmetry in the vicinity of
ring~1 in HD~163296. Such a feature arises naturally in planet-disk
interaction models including dust in the form of dust trapping in a Lagrange
point of the gap carving planet. In this case planet~1 is responsible for
dust trapping in the trailing L5 point which is also visible in
Fig.~\ref{fig:dust_hr} and Fig.~\ref{fig:dust_alpha} for a significant subset
of the parameter space. An equivalent result was presented in
\cite{Isella2018}. In the following subsections we aim to perform a more
extensive analysis of this feature in order to constrain physical properties
of the dynamical system.\\

\subsubsection{Structure \& dust feedback} \label{sec:structure_feedback}
In Fig.~\ref{fig:asymmetry_polar} in the left panels dust surface density
maps are shown in polar coordinates for four different dust fluids of model
\texttt{fid\_dres}. The region is focused around the co-orbital region of planet~1.
Clearly, dust is concentrated in the trailing Lagrange point L5 of the
Jupiter mass planet at 48~au. Several trends become apparent: Not
surprisingly, dust grains are trapped more efficiently for larger grain sizes \revthird{ and Stokes numbers}
due to the stronger drift. Furthermore, the shape is more elongated for
\revthird{ smaller Stokes numbers.} \\ Does the dynamics of this feature
change with
the consideration of a dust back-reaction onto the gas? An example of the
impact from dust feedback (model \texttt{p1m1fb\_dres}) is given in
Fig.~\ref{fig:asymmetry_polar} on the right hand side with the same
parameters as model \texttt{fid\_dres}. 
\revised{
The \revthird{crescent-shaped feature} exhibits different structures
compared to model \texttt{fid\_dres} at the location of L5. The dust
back-reaction onto the gas triggers and instability leading to fragmentation
of the dust feature.
Unlike the fiducial model, the additional crescent in
the leading Lagrange point L4 of planet~1 is more pronounced than the L5 feature for smaller grain sizes and Stokes
numbers. In the explored parameter space dust is significantly trapped for
small aspect ratios or very low values of $\alpha < 10^{-4}$ without the
modeling of dust feedback. With the fiducial set of parameters the L4 feature slowly dissipates after more than a 1000 orbits. 
\\ A closer look on the radial and azimuthal extent of these dust
shapes is provided in Fig.~\ref{fig:asymmetry_width} for all simulated dust
fluids. The surface densities of the \revthird{crescent-shaped asymmetry} are normalized to its maximum
value. The radial and azimuthal width increases with decreasing values of the
\revthird{Stokes number}. In the lower panel the density peak at the trailing L5 dominates
the leading peak at L4 for all dust species. For smaller \revthird{Stokes numbers and} grain sizes the
density maximum moves away from the planet location similar to the peak shift
of the concentric dust rings described in Sec.~\ref{sec:rings}.\\ In
Fig.~\ref{fig:asymmetry_width_fb} we display the results of model
\texttt{p1m6fb\_dres}. 
\revthird{
Contrary to the nonfeedback case the density peak at
L4 becomes significant for $\mathrm{St} \leq 3.6 \cdot 10^{-2}$.
}
\revsecond{
In the vicinity of the L5 point two density
peaks appear due to the fragmentation by dust feedback.\\
The azimuthal gas density profile reveals the momentary location of the
spiral wakes caused by the planets as well as the gas accumulation around
planet~1 itself at $\phi = 0$.}
}

\subsubsection{Dynamical stability}
In principle the \revthird{crescent-shaped features} in the co-orbital region are subject to diffusive
processes like dust diffusion due to turbulent mixing or gravitational
interaction with the planetary system (e.g. eccentric orbits). The dust
trapping mechanism has to counteract these disruptive forces for the feature
to be dynamically stable.\\ Fig.~\ref{fig:asymmetry_time_alpha} corroborates
this line of argument. The stability of the L5 feature is sensitive to the
local value of the $\alpha$ viscosity. Given a value of $\alpha = 10^{-5}$
the feature remains stable throughout almost the entirety of the simulation
whereas $\alpha = 10^{-3}$ shortens the existence down to about 300 dynamical
time scales. For larger viscosities no discernable feature develops and the
co-orbital region simply empties its dust content from the initial
condition.\\ Studying the results of the models \texttt{p1m1} to
\texttt{p1m5} we find that below about 0.4 to 0.5 Jupiter masses for planet~1
no stable feature forms since the gravitational interaction is not sufficient
to enforce dust trapping in the Lagrange points. Qualitatively the feature
life time is thus very sensitive to the given physical parameters. It should
be noted that the absolute value in dynamical time scales could be
underestimated due to the numerical diffusion present in the lower resolution runs.
This additional diffusive effect prevents a stable trapping
region around L4 and L5. Further details are provided in
appendix~\ref{sec:appendix_res_study}.
\revised{
In Fig. \ref{fig:asymmetry_time_alpha} we therefore plotted results with 14 cells per scale height.
}

\subsubsection{Dust mass}
A substantial amount of dust can be trapped in the feature at L5. The total
mass ranges from roughly $1 \,M_\mathrm{earth}$ for the low mass model to
$10-15 \,M_\mathrm{earth}$ for the high mass model. An overview of relevant
parameters and their impact on the trapped dust mass is given in
Fig.~\ref{fig:plateau_mass_evolution}. Generally, if a sufficiently stable
\revthird{crescent-shaped asymmetry} develops, the order of magnitude of the trapped dust mass is
comparable for all parameters. Here, we only consider the three largest dust
species since smaller grains are prone to be weakly trapped.
\\ 
\revthird{Only the low resolution simulations are compared to each other in this parameter study.
The mass of the crescent-shaped feature was averaged over 200 orbits starting from $100 \, T_0$ when convergence is reached. }
Looking at the
aspect ratio dependence, we find that an increase in $H / r$ also leads to a
higher dust mass of the L5 feature. A more massive planet also causes an
increase in the trapped dust mass. The lighter the planet, the less stable
the agglomeration of dust becomes. As mentioned above, for less than about
$0.4$ to $0.5\,M_\mathrm{jup}$ no stable feature forms at L5. Considering the
influence of the $\alpha$ viscosity parameter a local value of $\alpha \geq
10^{-3}$ causes a significant loss of dust mass. For $\alpha \geq 2\cdot
10^{-3}$ no feature develops. \revthird{Simulation results with a radially constant value of $\alpha$ are used here} \\ 
Finally, the introduction of an eccentric planetary orbit
leads to an almost linear decrease of the trapped dust mass with respect to
the eccentricity value. 
\revthird{
For values $\geq 0.06$ no feature forms for dust Stokes numbers of $3.6 \cdot 10^{-2}$ and below. }
\\
\revised{
  An increase of the mass of planet~2 has a small influence on the trapped
  dust mass. The continuous gravitational interaction perturbs the \revthird{crescent-shaped asymmetry}
  decreases the amount of mass trapped in the feature. The difference between
  $0.3 M_\mathrm{jup}$ and $0.9 M_\mathrm{jup}$ however accounts to roughly
  5\% of trapped dust mass.
}

\subsubsection{Growth time scale}
\revised {
The growth time scale of the planet mass can have a significant impact on the
formation of vortices \citep{hammer_2017,hammer_2019,hallam_2020}. Therefore, simulation
runs including longer growth time scales with values of $T_\mathrm{G} = 10$,
50, 100 and 500 orbits were performed. Qualitatively, the results are mostly
unaffected by the choice of $T_\mathrm{G}$. In
Fig.~\ref{fig:plateau_mass_evolution} the masses only deviate about 10 \%
from the fiducial model. With the longest growth time scale of 500 orbits
smaller grains are trapped more efficiently while the mass contained in the
largest grains decreases slightly. Deviations from the fiducial model for
these long growth time scales could also be caused by a loss of dust content
and local redistribution of grain sizes due to dust drift in the simulation
domain. More details are given in Appendix \ref{sec:appendix_growth_timescale}.
}
\\
\revsecond{The formation of vortices is sensitive to the planet growth time
scale since the vortex smooths out the gap edge and reduces the steepness of
the corresponding edge slope which in turn weakens the Rossby wave
instability \citep{hammer_2017}. This effect is important for longer planet
growth time scales and leads to weaker, elongated vortices. In the context of
dust trapping in the Lagrange points however, the process takes place in the
co-orbital region and the amount of mass concentrating in the asymmetries is
determined by the initial dust content available within this region
\citep{montesinos_2020}. Since the dust trapping here is related to the
horseshoe motion in the co-orbital region, it is a different mechanism and
the evolution of the gap edge does not seem to have a major influence on the
dynamical origin of the asymmetric features. }

\subsection{Synthetic images} \label{sec:synth_images}
The main question remains if dust trapping in the L5 point in the models
presented could explain the observed feature in HD~163296. We apply the
procedure described in Sec.~\ref{sec:radtrans} and thus extend the surface
density maps to three dimensional grids, perform dust radiative transfer
calculations RADMC-3D and simulate the observation with ALMA by using the
CASA package accordingly. 
\revthird{
Snapshots of the density maps for these synthetic images are shown in
Fig.~\ref{fig:dust_overview}. Both the high resolution fiducial model
\texttt{fid\_dres} and the dust feedback model \texttt{p1m6fb\_dres} are
used. 
}
\\ The resulting dust density grid crucially
depends on the vertical dust scale height $H_\mathrm{d}$ and thus the dust
settling prescription (see Eq.~\ref{eq:dust_settling}). With the assumption
of constant grain sizes throughout the disk the local Stokes number is used
for the calculation of $H_\mathrm{d}$.
\revthird{The grain sizes distribution is the same as the one in the
simulation runs. Depending on the low or high mass model, the grain sizes and
opacities and densities were adjusted accordingly.}
We chose to leave $\alpha$ as a free
parameter for the vertical dust settling recipe which will be denoted as
$\alpha_\mathrm{z}$. 
\revsecond{
In Fig.~\ref{fig:synth_images} synthetic images from various snapshots of the simulation are
compared to the observation. The images qualitatively reproduce the observed
features. The key difference with respect to the crescent-shaped feature is the
more elongated shape compared to the fiducial model. Furthermore, the models
produce a radially symmetrically located feature while the observed one is
situated closer to ring~1. After 500 and 1000 orbits at 48 au the feature at
L4 is still visible due to the slow dissipation at the L4 point. In the image
computed from the snapshot at 2000 orbits of the simulation \texttt{fid\_dres}, only the L5 feature
appears, as also seen in Fig.~\ref{fig:dust_overview}. \\
As already
discussed in Sec.~\ref{sec:structure_feedback} and
Fig.~\ref{fig:asymmetry_polar} a significant amount of dust agglomerates in
the L4 region for smaller grain sizes and Stokes numbers. This effect is
clearly visible in the synthetic image of \texttt{p1m6fb\_dres} at 500 orbits with dust
feedback enabled in Fig.~\ref{fig:synth_images}. No such feature is present
in the observation.}
The fragmentation of the \revthird{crescent-shaped asymmetry} around L5 becomes more apparent in the
later stages of the simulation. After 1000 orbits dust is concentrated in
clumps of small azimuthal extent. These features are substantially different
from the observation.
\revthird{Since the high mass model assumes a lower dust-to-gas ratio, the
synthetic images created from the high mass model and the respective grain
size distribution only serve as a comparison of the visibility of such
features in the co-orbital region. Additionally, synthetic observations based
on the low mass model are shown in Fig.~\ref{fig:synth_images}. As expected,
the larger values of the opacity for the grains with the maximum Stokes
numbers compared to the high mass model leads to much narrower rings and
substructures. The high mass model is thus more suitable for the HD~163296
system.
}
\\ Focusing on ring~2, the intensity is slightly
lower compared to the observations. This result is consistent with
Fig.~\ref{fig:surface_dens_tau} where the optical depth of ring~2 does not
reach the derived values of \cite{Huang2018a} for smaller grains. \\ Looking at the inner part
of the disk within the gap at 48~au, the simulated images show a secondary
gap caused by planet~1 which is not present or visible in the observed
structure.
\revsecond{
Furthermore, the influence of vertical mixing of dust grains can be
investigated with the three synthetic images in Fig.~\ref{fig:synth_images_alpha} of the nonfeedback model. 
}
The model run with an increased dust scale height ($\alpha_\mathrm{z} =
10^{-4}$) displays a more diffuse intensity map and a slight decrease in
intensity perpendicular to the axis of inclination. A difference in ring
thickness depending on the azimuthal location is not visible in the observed
system. For the weakest vertical mixing ($\alpha_\mathrm{z} = 10^{-6}$) the
dust substructures appear completely flat. Ring~2 is much fainter than the
observed intensity and ring~1 appears significantly thinner. The model with
$\alpha_\mathrm{z} = 10^{-5}$ comes closer to the observed ring thickness
while having a mostly azimuthally constant ring structure.\\
Assuming dust trapping in the L5 point of the observation we can propose
potential coordinates for a yet undetected planet. Comparing the results of
model \texttt{fid\_dres} with the ALMA image, the planet offset relative to the
disk center is $\delta\mathrm{RA} \approx -0.352 \, \mathrm{arcsec}$ and
$\delta\mathrm{DEC} \approx 0.104 \, \mathrm{arcsec}$. This corresponds to
the coordinates \texttt{RA=17h56m21.2563s, DEC=-21d57m22.3795s}.

\section{Discussion} \label{sec:discussion}
In the following parts we compare our results with previous works on the
HD~163296 system and equivalent simulations as well as limits and caveats of
the models presented here.

\subsection{Comparison to previous works}
Studying the observed gap widths \cite{Isella2016} postulated a range of
$0.5\, M_\mathrm{jup}$ to $2\,M_\mathrm{jup}$ for planet~1 at 48~au,
$0.05\,M_\mathrm{jup}$ to $0.3\,M_\mathrm{jup}$ for planet~2 at 83~au and
$0.15\,M_\mathrm{jup}$ to $0.5 \,M_\mathrm{jup}$ for planet~3 at 137~au. With
more detailed hydrodynamical models by \cite{Liu2018} using a multi-fluid
dust approach the planet masses were constrained to $0.46\,M_\mathrm{jup},
0.46\,M_\mathrm{jup}$ and $0.58\,M_\mathrm{jup}$ for the three planets
respectively. At this point no asymmetries were observationally resolved. \\
Among the publication of the DSHARP survey \cite{Zhang2018} performed an
extensive parameter study with hydrodynamical planet-disk interaction
simulations using a Lagrangian particle dust formalism. Their results
indicate planet masses of $0.35\,M_\mathrm{jup}, 1.07\,M_\mathrm{jup}$ and
$0.07\,M_\mathrm{jup}$ if a radially constant $\alpha$ viscosity of $10^{-4}$
is assumed. The predicted masses of $1\,M_\mathrm{jup}$ and $1.3\,
M_\mathrm{jup}$ by \cite{Teague2018} for the two outer planets exceed the
hydrodynamical results. However, with the uncertainties of about $50 \%$ the
planet masses used in the models can be consistent with the kinematical
detections. \\ Our models indicate that for fiducial model parameters, e.g.
an aspect ratio of 0.05 and a radially increasing $\alpha$ viscosity similar
to \cite{Liu2018}, a minimum mass of $\approx 0.5\,M_\mathrm{jup}$ for
planet~1 is necessary to produce a stable dust trap in the trailing Lagrange
point L5. For higher masses, the amount of dust trapped in the
\revthird{crescent-shaped asymmetry} can be slightly decreased
(see~\ref{sec:appendix_planet2}). \\ The initial gas surface density at
48~au of $\Sigma_\mathrm{g, 0} = 37.4 \mathrm{g} / \mathrm{cm}^2$ for the
high mass model assuming a local dust-to-gas mass ratio of $\approx 2.4 \cdot
10^{-3}$ is close to the findings of \cite{Zhang2018} with $\Sigma_\mathrm{g,
0} = 3 - 30 \mathrm{g} / \mathrm{cm}^2$. \cite{Isella2016} used a value of
$\Sigma_\mathrm{g, 0} = 10 \mathrm{g} / \mathrm{cm}^2$. Given the proximity
of the \revthird{crescent-shaped asymmetry} and ring~1 in the observations, it is a natural choice to
normalize the dust density to the values derived from the optical depth
comparison. \\
\revsecond{
\cite{marzari_1998} found that if planetesimals are small enough to be
affected by the gas drag, the stability of the L4 point is reduced and the
density distribution of L4 and L5 becomes asymmetric. A similar effect was
observed in the gas by \cite{masset_2002} if viscosity is included. In this
case, compared to the gas drag affecting the dust, the viscous gas drag acts
as the effect causing the asymmetric gas distribution. Similar results in the
context of hydrodynamical simulations including gas and dust were found by
\cite{lyra_2009}. Recently, \cite{montesinos_2020} presented hydrodynamical
simulations including multi-species particle dust exploring the stability of
the L4 and L5 in the presence of a massive planet with at least one Jupiter
mass. Their findings basically agree with the results presented in this paper
without the effect of dust feedback. They state, that L5 captures a larger
amount of dust compared to the L4 point. They argue that colder disks allow
for more efficient dust trapping in these Lagrange points, lower viscosity
leads to a more symmetric distribution of dust in L4 and L5 and dust entering
the co-orbital region from the outer part of the disks seems to not
significantly contribute to the mass of the clumps in L4 and L5. \\
Interestingly, the Trojans populating L4 and L5 around Jupiter seem to be
more numerous around the L4 point \citep{yoshida_2005}. In our models this
effect only appears if dust feedback plays a significant role in this region.
}

\subsection{Model assumptions}
Dust opacities are highly sensitive to its material composition and spatial
structure. Estimating the surface densities from the optical depth is thus
subject to a significant uncertainty. This is amplified by the choice of the
dust size distribution and dust size limits. However, the features of
interest, i.e. the \revthird{crescent-shaped feature} and ring~1 match the observations
reasonably well with the high mass model, \revthird{setting $a_\mathrm{min} = 0.19 \,
\mathrm{mm}$ and $a_\mathrm{max} = 19 \,\mathrm{mm}$ with the MRN size
distribution.}\\ The life time of the \revthird{crescent-shaped asymmetry} in L5 depends on diffusive
processes like the turbulent viscosity and mixing of dust grains. In the
resolution study (see appendix~\ref{sec:appendix_res_study}) the resulting
life time seems not to be limited within the simulated time frame. In lower
resolution studies investigated in this paper the numeric diffusion artificially
truncates the feature's life time. Even by employing highly resolved
simulations the age estimation of the \revthird{crescent-shaped feature} and thus approximately the
planet itself would be difficult with the degenerate parameter space.
Eccentricity and viscosity both shorten the time scale of dispersal
significantly. More detailed studies and observations are necessary to
constrain dynamical age of the substructures which need to be performed at
higher spacial resolution.\\ In the observation presented in
\cite{Isella2018} the \revthird{crescent-shaped asymmetry} is located at $r = 55\, \mathrm{au}$ instead
of $r = 48 \,\mathrm{au}$. No combination of parameters in our models are
able to reproduce this effect. An eccentric planet would be an intuitive
choice but only leads to a disruption of the \revsecond{crescent-shaped feature}. Dust
feedback can lead to an unstable feature, ultimately leading to small clumps.
In the earlier stages, dust feedback promotes dust
trapping in the L4 point. No such effect is seen in the observations. Another
explanation of the positioning of the \revthird{crescent-shaped asymmetry} could be planet migration.
Depending on the migration direction and speed, the locations of the rings
and features in the co-rotation region can be asymmetrically shifted in
radial direction \citep{meru_2018, perez_2019, Weber2019}. Additionally,
sudden migration jumps in a system of multiple planets can temporarily create
trailing asymmetries with respect to the migrating planet as shown in
Rometsch et al. (submitted). \\ It should be noted that dust coagulation and
fragmentation is not considered here. More sophisticated models including
these effects as shown in \cite{drazkowska_2019} could be used in this case
but are computationally demanding.\\ Ring~2 is slightly fainter in our
models compared to the observations. The amount of dust that can be trapped
in ring~2 depends on planet~3 since it truncates the dust flow from the outer
part of the disk. One hypothesis might be that planet~3 formed later than
planet~2 and thus allowed a larger amount of dust to be accumulated in the
second ring. As shown in Fig.~\ref{fig:dust_alpha} it is furthermore possible
to confine the range of permissible values of $\alpha$ by the disappearance
of ring~2 for large viscosities ($\alpha > 2\cdot 10^{-3}$) and low
viscosities ($\alpha < 5\cdot10^{-4}$) due to vortex activity.\\ The
synthetic images in Fig.~\ref{fig:synth_images} display an additional gap in
the inner dust disk. This secondary is caused by the interaction with the
spiral wakes originating from planet~1. The effect is mostly visible for
large Stokes numbers and dust sizes as well as high planet masses. However, a
close to Jupiter mass planet is necessary to trap the needed amount of dust
in the L5 point to be comparable to the observations. Results of
\cite{miranda_rafikov_2019} indicate that radiative effects are important,
even at large distances of the central star, since locally isothermal models
over-pronounce the effect of the spiral wakes and secondary gaps. The same
effect was shown to be important for the inner gas disk of HD~163296 in the
work of \cite{ziampras_2020}. It can be expected that the additional
secondary ring in the inner disk disappears when radiative effects are taken
into account. Nevertheless, the inner dust disk is not the main aspect of our
work and the locally isothermal approach can be considered to be sufficient
for modeling the \revthird{crescent-shaped feature}. \\
\revised{
  The planet growth time scale has only a minor impact on the overall dust
  substructure emerging in the simulations. Differences are likely caused by
  the change in dust content and local dust size distribution due to dust
  drift. Longer growth time scales lead to a lower intensity of ring~2 due to
  the lack of material that has already drifted inwards before being trapped
  by the outer planets. The dynamical structure, especially the shape and
  location of the \revthird{crescent-shaped asymmetry}, is basically unaffected within the explored
  parameter space of growth time scales.\\
}
\revthird{
  The synthetic images based on the low mass model show narrower rings and
  differ significantly from the observations. The high mass model is thus
  favored in this study.
}

\section{Conclusion} 
\label{sec:conclusion}
We presented a parameter study of the \revthird{crescent-shaped feature} of the
protoplanetary disk around HD~163296 using multi-fluid hydrodynamical
simulations with the FARGO3D code. The model includes eight dust fluids with
initial Stokes numbers ranging from $\mathrm{St}=1.3\cdot10^{-3}$ to
$\mathrm{St}=1.3\cdot10^{-1}$ and grain sizes of $a_\mathrm{min} = 0.19 \,
\mathrm{mm}$ and $a_\mathrm{max} = 19 \, \mathrm{mm}$ for the high mass model. Additionally,
synthetic ALMA observations based on radiative transfer models of the
hydrodynamical outputs are presented. Comparing the model with the
observation, the results match qualitatively.\\ In this work we showed that
the observation of the \revthird{crescent-shaped feature} puts important
constraint on the disk and planet parameters -- always under the assumption
that the feature is truly caused by dust accumulation in the planet's
trailing Lagrange point L5. Most importantly, it confines the level of
viscosity and planetary mass. The main findings can be summarized as follows:
\begin{enumerate}
    \item The observed \revthird{crescent-shaped asymmetry} in the observation \citep{Isella2018} can be
    reproduced with a Jupiter mass planet in the respective gap location at
    48~au. Dust is effectively trapped in the trailing Lagrange point L5. In
    the case of negligible dust feedback the L4 point is not sufficiently
    populated to be observable. The peak of the asymmetric dust density
    distribution shifts towards the planet location for larger Stokes numbers
    and grain sizes.
    \item Rescaling the dust densities to the observed optical depth of
    ring~1 at 67~au dust masses of 10 to 15 earth masses can be trapped in a
    crescent shaped feature located at the L5 point. The trapped dust mass is
    relatively insensitive to the choice of viscosity, aspect ratio, planet
    mass and eccentricity as well as the planet growth time scale.
    \item Including the dust back reaction onto the gas can lead to dust
    trapping preferably at the leading Lagrange point L4 \revthird{ for initial Stoke numbers of $\mathrm{St} \leq
    3.6\cdot10^{-2}$ and at later stages to fragmentation of the crescent-shaped asymmetry}
    near the L5 point.
    \item Diffusive and disruptive effects counter the stability of the dust
    trap in L5. Values of $\alpha \geq 2\cdot10^{-3}$ prevent the formation
    of an asymmetric and stable feature. Introducing eccentricity leads to
    the same result. The shifted location of the observed \revthird{crescent-shaped feature} at 55~au
    is not justified by an eccentric planet carving the corresponding gap in
    the given parameter space.
    \item If the L5 feature is caused by an embedded planet, the models allow
    an estimation of the azimuthal planet position in the gap. The planet
    offset relative to the disk center is $\delta\mathrm{RA} \approx -0.352
    \, \mathrm{arcsec}$ and $\delta\mathrm{DEC} \approx 0.104 \,
    \mathrm{arcsec}$ which corresponds to the coordinates
    \texttt{RA=17h56m21.2563s, DEC=-21d57m22.3795s}.
\end{enumerate}
We can thus conclude that a combination of $\approx 1 M_\mathrm{jup}$ and
$\approx 0.5 \,M_\mathrm{jup}$ for the inner planets in combination with a
MRN dust size distribution with $a_\mathrm{min} = 0.19 \,\mathrm{mm}$ and
$a_\mathrm{max} = 19\,\mathrm{mm}$ as well as a local value of $\alpha = 2
\cdot 10^{-4}$ can reproduce the observed \revthird{crescent-shaped asymmetry} and ring structures
sufficiently well.
\revised{ The dust-to-gas ratio in the models may be overestimated
since none of the features emerging in the simulations including feedback,
e.g. two \revthird{crescent-shaped asymmetries} and fragmentation, are present in the observation.
Additional high resolution studies are necessary to constrain the parameter
space further, also in regard to the long-term stability of the feature.
}
\begin{acknowledgements}
  Authors Rodenkirch, Rometsch, Dullemond and Kley acknowledge funding from
  the DFG research group FOR 2634 ''Planet Formation Witnesses and Probes:
  Transition Disks'' under grant DU 414/23-1 and KL 650/29-1, 650/30-1.
  The research leading to these results
  has received funding from the European Research Council under the European
  Union’s Horizon 2020 research and innovation programme (grant agreement No.
  638596; P.W.). The authors acknowledge support by the
  High Performance and Cloud Computing Group at the Zentrum f\"ur
  Datenverarbeitung of the University of T\"ubingen, the state of
  Baden-W\"urttemberg through bwHPC and the German Research Foundation (DFG)
  through grant INST\,37/935-1\,FUGG.
  Plots in this paper were made with the Python library \texttt{matplotlib} \citep{hunter-2007}.
\end{acknowledgements}

\bibliographystyle{aa}
\bibliography{main}

\appendix

\section{Secondary planet mass} \label{sec:appendix_planet2}
\begin{figure*}[!t] 
  \centering
  \includegraphics[width=\linewidth]{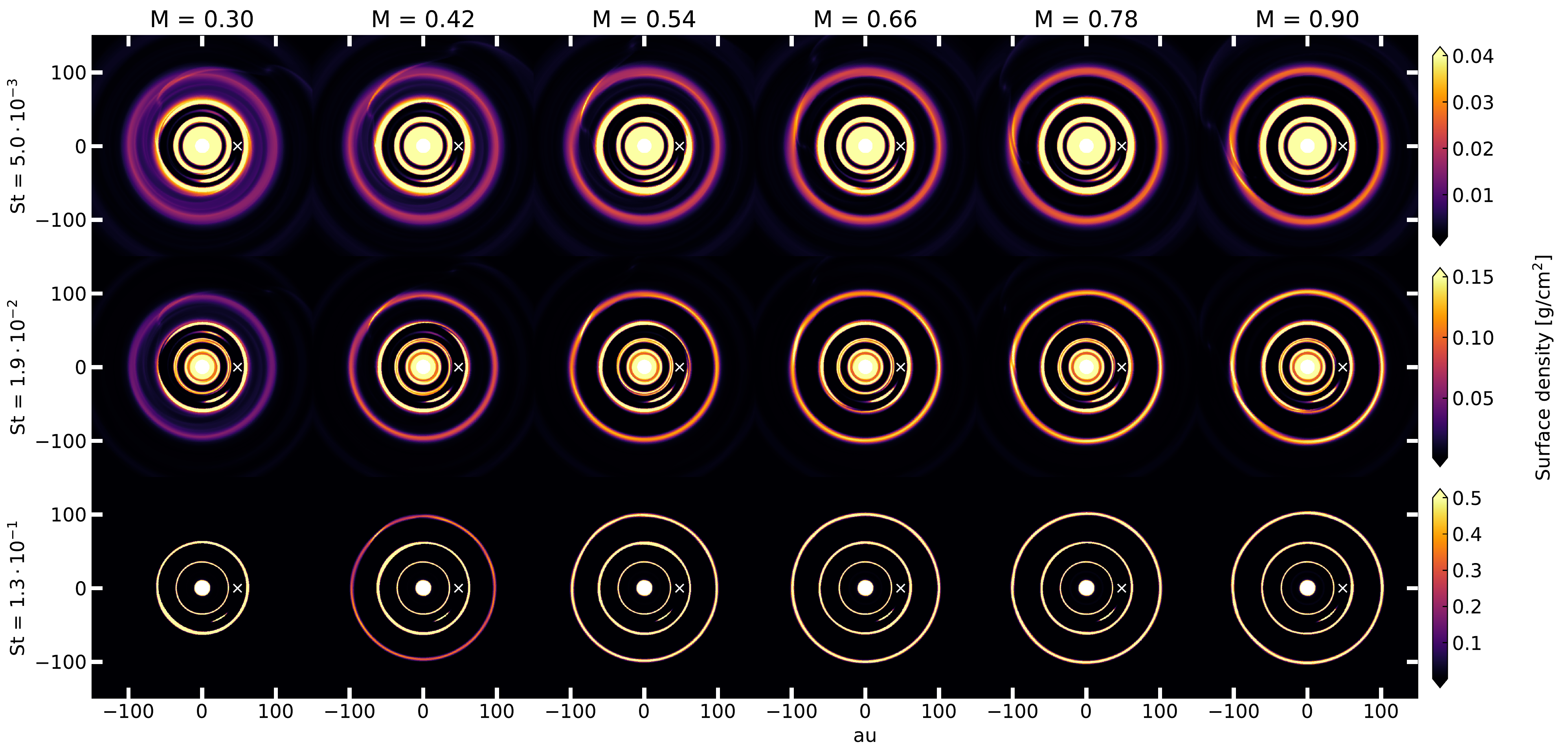}
  \caption{Dust surface density maps for a subset of three fluids with
  varying values of the planet~2 mass in $M_\mathrm{jup}$ at 500 orbits at
  $r_0$.
  \revsecond{The white crosses mark the position of planet~1.}
  }
  \label{fig:dust_planet2}
\end{figure*}

In Fig.~\ref{fig:dust_planet2} a parameter study of the planet~2 mass
influence is shown, involving the models \texttt{p2m1} to \texttt{p2m6}.
The nonaxisymmetric feature at the L5 point is
sensitive to the mass of planet~2. In general, the passing of the planet acts
as a perturber, inhibiting an effective dust trap in L5. The effect is
visible in Fig.~\ref{fig:dust_planet2} for smaller grain sizes. Lower planet
masses weaken the dust trapping in ring~2. We identify the balance between
effective trapping in ring~2 and optimal dust trapping in the \revthird{crescent-shaped feature} to be on the order of
$\approx 0.5 M_\mathrm{jup}$, thus the choice of the fiducial model
parameter.

\section{Resolution study} \label{sec:appendix_res_study}
\begin{figure}[htb] 
  \centering
  \includegraphics[width=\linewidth]{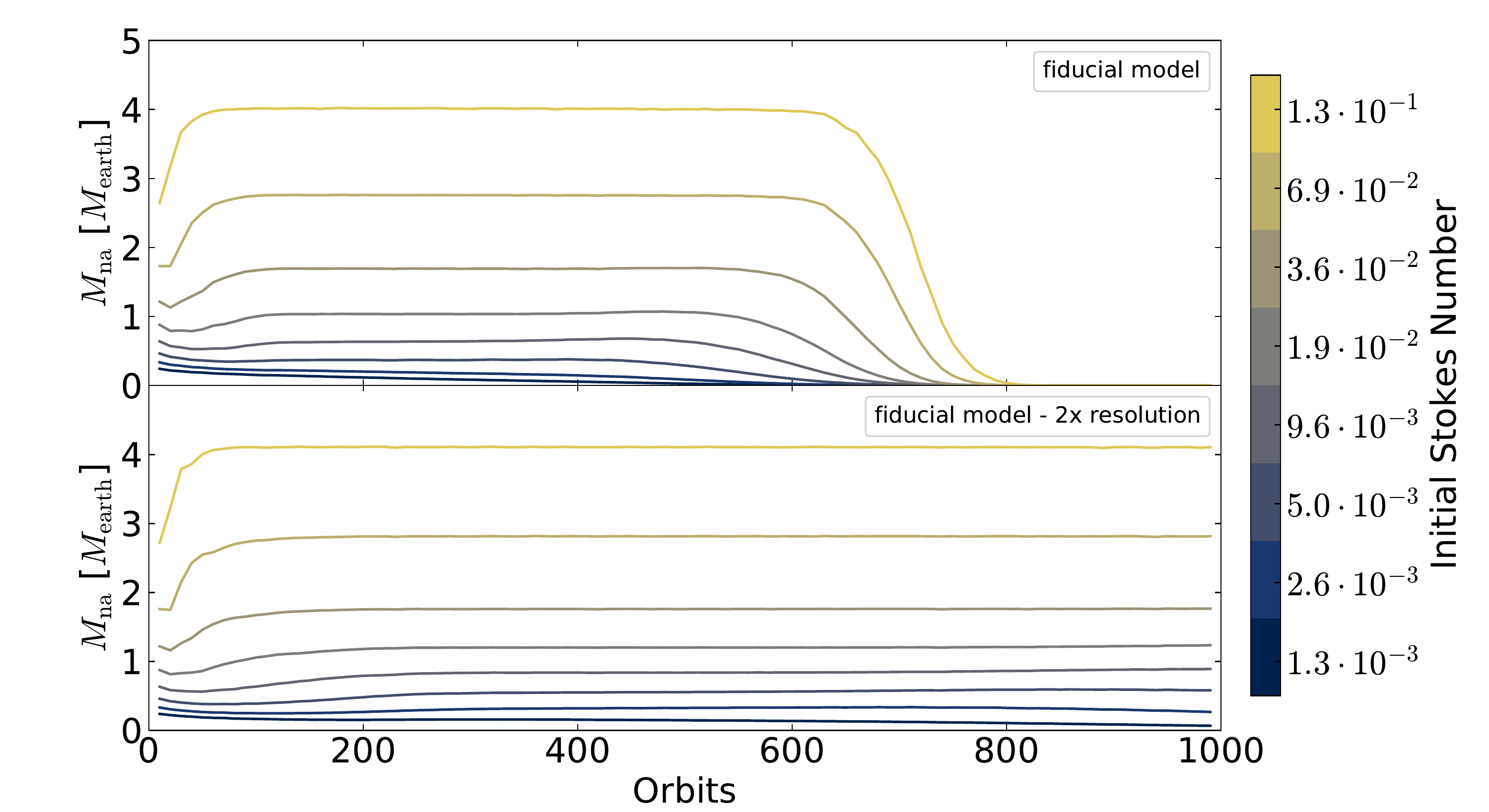}
  \caption{Development of the trapped dust mass in the L5 region of the
  simulations \texttt{fid} and \texttt{fid\_dres} over time. In the
  lower panel the grid resolution is doubled in both the radial and azimuthal
  direction. The dispersal of the features are prolonged in the high resolution
  setup.}
  \label{fig:res_study}
\end{figure}

In the model \texttt{fid\_dres} the resolution is doubled to $1120 \times 1790$ cells in radial and azimuthal direction
respectively compared to model
\texttt{fid}. Fig.~\ref{fig:res_study} indicates that an increased
resolution leads to a more stable asymmetry in the Lagrange point L5. No
decline in mass can be observed in the simulated time frame (1000 orbits at
48~au). The feature in the low resolution model \texttt{fid} however
depletes rapidly after 600 orbits. \\ Since the dust trapped
in this feature is sensitive to diffusive and disruptive effects, like e.g.
viscosity, eccentricity and the passing of the outer planet, the accelerated
dispersal may be attributed to numerical diffusion. \\
\revsecond{
Quantifying the absolute value of the numerical diffusion is complex, however
the order of magnitude can be estimated by a simple comparison of the high
resolution run \texttt{alpha3\_dres} with $\alpha = 5 \cdot 10^{-4}$ and the
fiducial model. As shown in Fig.~\ref{fig:asymmetry_time_alpha} the feature
lifetime is approximately comparable to the one of the lower resolution run
\texttt{fid} with a local $\alpha = 2\cdot 10^{-4}$ . Therefore, the effect
of the numerical diffusion in the low resolution model should be
approximately equivalent to $\alpha \approx 5 \cdot 10^{-4}$. Since FARGO3D
is second-order accurate in space and the error of the dust module has been
found to be proportional to a power law with an exponent of -2.2 as a
function of the number of grid cells \citep{Benitez-llambay2019}, the
numerical diffusion is expected to be equivalent to $\alpha \approx 1 \cdot
10^{-4}$ in the model \texttt{fid\_dres}. The resolution is thus sufficient
to describe the effect of prescribed local viscosity of $\alpha = 2\cdot
10^{-4}$ . \\ Nevertheless, the absolute values of the amount of trapped dust
mass in the stable phase is not significantly affected by the low resolution
effect and lower resolution models are thus acceptable to quantify these
values.
}

\section{Planet growth time scale} 
\label{sec:appendix_growth_timescale}

\begin{figure*}[!t] 
  \centering
  \includegraphics[width=\linewidth]{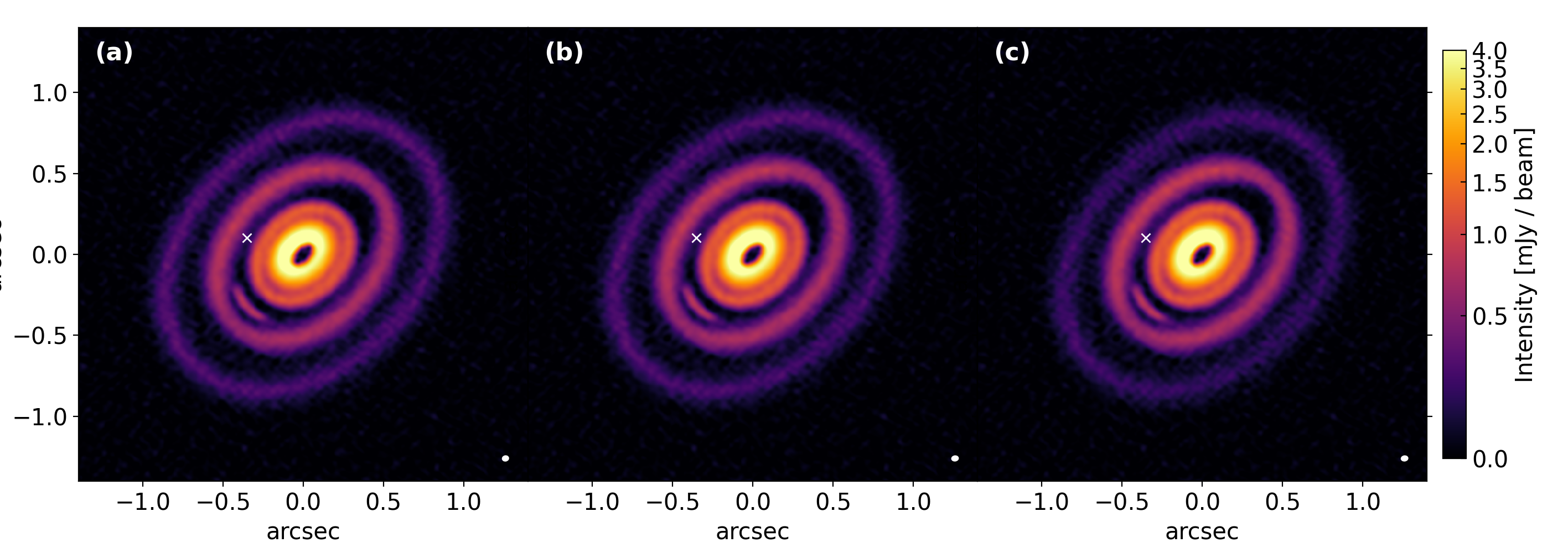}
  \caption{\revised{Comparison of the synthetic images of different planet growth time scales (a):
  $T_\mathrm{G} = 10\,T_0$, (b): $T_\mathrm{G} = 100\,T_0$, (c):
  $T_\mathrm{G} = 500\,T_0$. The snapshots are taken after 2000 orbits at 48 au.}}
  \label{fig:synthetic_taper}
\end{figure*}

\begin{figure}[htb] 
  \centering
  \includegraphics[width=\linewidth]{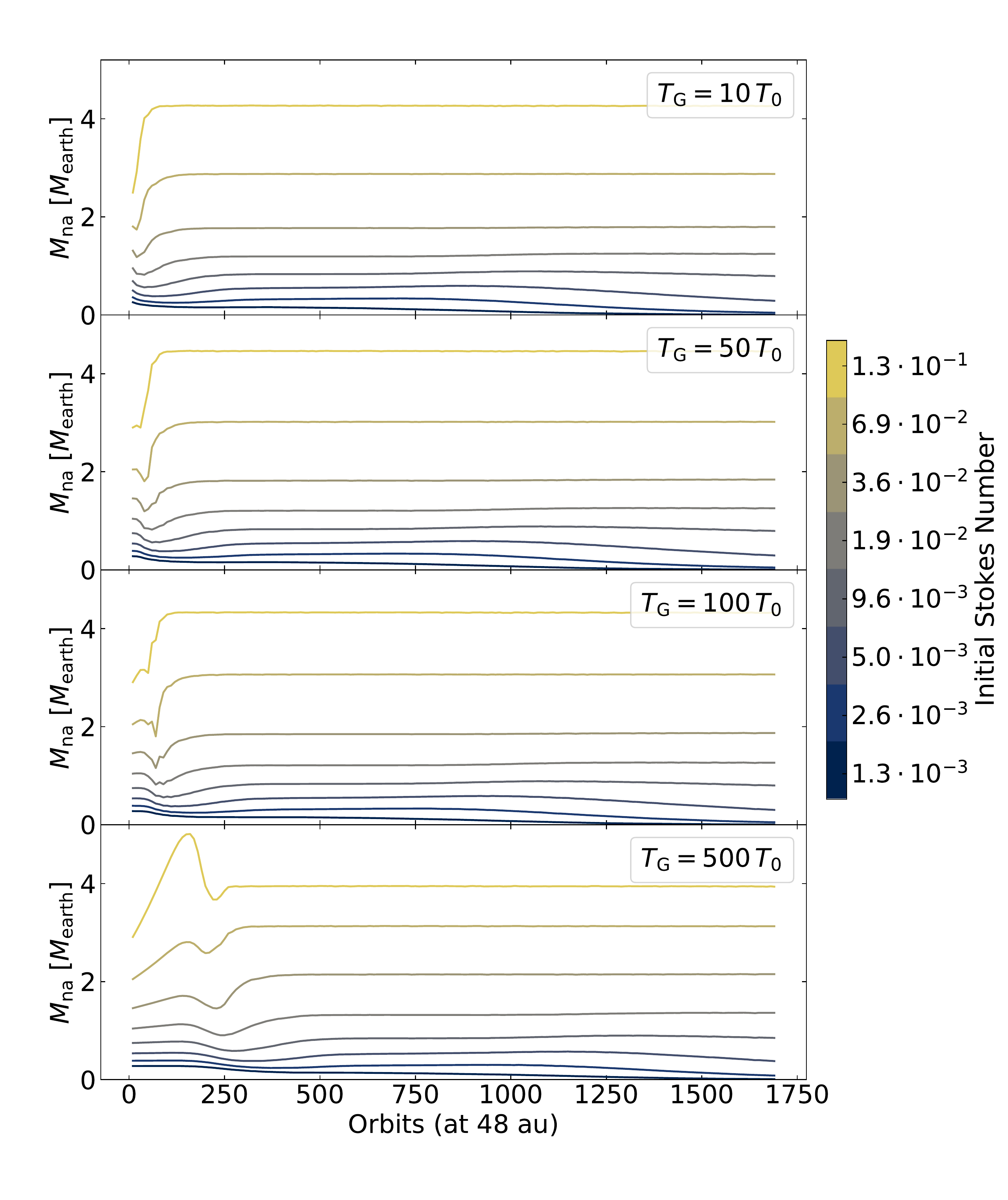}
  \caption{\revised{Evolution of the trapped dust mass $M_\mathrm{na}$ in the
  asymmetry around the L5 point for different planet growth time scales
  $T_\mathrm{G}$.}}
  \label{fig:time_taper}
\end{figure}

\revised{
  In Fig. \ref{fig:synthetic_taper} simulated ALMA observations are shown for
  planet mass growth time scales ranging from $T_\mathrm{G} = 10\, T_0$ to
  $500\, T_0$. Ring~2 becomes less massive for longer planet growth
  time scales since dust drift depletes the outer regions before the planets
  reach a sufficiently high mass for efficient trapping. \\
  The inner disk structure including the \revthird{crescent-shaped asymmetry} remains unaffected by the
  choice of parameters.  Fig. \ref{fig:time_taper} reveals the temporal
  evolution of the dust content in the asymmetry around the L5 point for all
  four growth time scales. For all runs grains smaller than $\approx 1\,
  \mathrm{mm}$ become depleted after more than 1000 orbits of evolution.
  After initial jumps in dust mass all simulations reach a stable stationary
  state considering millimeter grain sizes and above.
}

\section{Dust temperatures} \label{sec:appendix_dust_temperatures}

\begin{figure}[htb] 
  \centering
  \includegraphics[width=\linewidth]{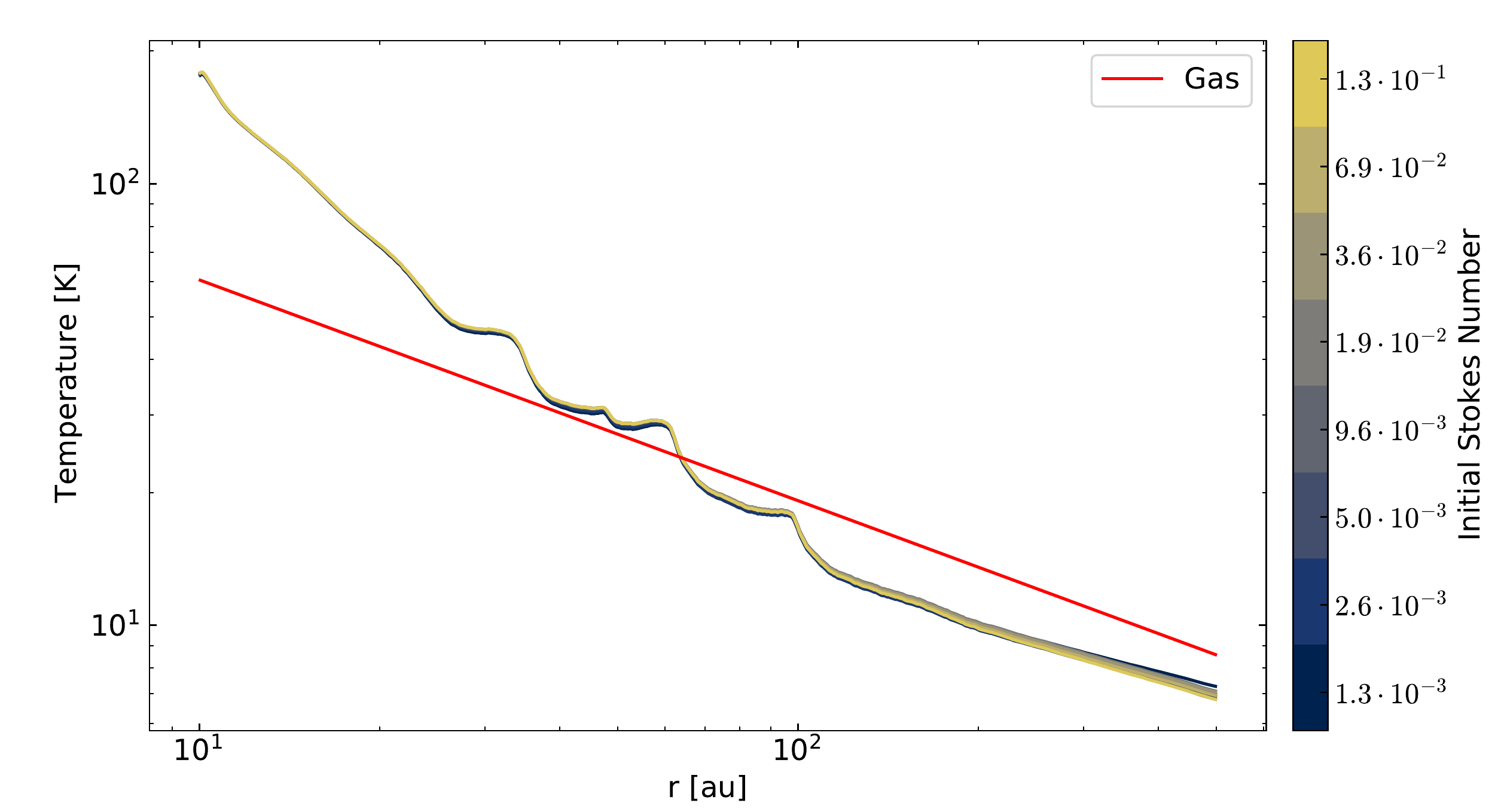}
  \caption{\revised{Comparison of prescribed gas temperature in the hydrodynamical
  simulation and the dust temperatures of the thermal Monte-Carlo calculation
  at the disk mid plane for all grain sizes.}}
  \label{fig:dust_temperatures}
\end{figure}

\revised{
In Fig. \ref{fig:dust_temperatures} dust temperatures from the radiative
transfer calculation of the fiducial model and the prescribed gas
temperatures are shown. While gas temperature gradient is smaller than the
one for the dust, the temperatures and the slope of both match well at the
location of planet~1, the primary region of interest where the \revthird{crescent-shaped asymmetry}
forms. For the large grain sizes studied in this paper, a gray body
approximation for the temperature is approximately valid. We thus see no
significant increase in temperature comparing the grain sizes to each other.
}

\end{document}